\documentstyle[aps,rmp,epsf,psfig]{revtex}
\tighten
\draft

\newcommand{\ab}{accelerator-based }
\newcommand{\rb}{reactor-based }
\newcommand{\nuebar}{$\bar{\nu}_e\;$}

\newcommand{\usec}{$\mu$s}

\begin{document}

\twocolumn[\hsize\textwidth\columnwidth\hsize\csname
@twocolumnfalse\endcsname

\title{Reactor-based Neutrino Oscillation Experiments \\}
\author{Carlo Bemporad$^1$, Giorgio Gratta$^2$, and Petr Vogel$^3$ \\ }
\address{$^1$INFN and University of Pisa, Pisa, Italy \\
$^2$Physics Department, Stanford University, Stanford CA 94305, USA \\
$^3$Physics Department, California Institute of Technology, Pasadena, CA
91125, USA \\}
\date{\today}

\maketitle

\begin{abstract}

\baselineskip 12pt
{\normalsize
 The status of neutrino oscillation searches
 employing nuclear reactors as sources is reviewed.  
This technique, a direct
 continuation of the experiments that proved the existence
 of neutrinos, is today an essential tool
 in investigating the indications of oscillations found in
 studying neutrinos produced in the  sun and in the earth's
 atmosphere.  The low-energy of the reactor \nuebar makes them
 an ideal tool to explore oscillations with small mass differences
 and relatively large mixing angles.
   In the last several years the determination of the reactor
 anti-neutrino flux and spectrum has reached a high degree of
 accuracy. Hence measurements of these quantities at a given
 distance $L$ can be readily compared with the expectation at $L = 0$,
 thus testing \nuebar  disappearance.
   While two experiments,
\textsc{Chooz}  and \textsc{Palo Verde}, 
with baselines of about 1 km and thus sensitive to the
 neutrino mass differences associated with the atmospheric neutrino
 anomaly, have collected data and published results recently,
 an ambitious project with a baseline of more than 100 km,
\textsc{Kamland}, is
 preparing to take data. This ultimate reactor experiment will
 have a sensitivity  sufficient to explore part of the oscillation phase
 space relevant to solar neutrino scenarios. It is the only envisioned
 experiment with a terrestrial source of neutrinos capable of addressing
 the solar neutrino puzzle.}

\end{abstract}

\vspace{0.5cm}]


\newpage

\narrowtext

\section{Introduction}

Neutrinos have the  distinction
of being the first elementary particle whose existence was 
predicted by a theorist in order to explain seemingly
unrelated phenomena\footnote
{For early development in neutrino physics see,
for example,  Chapter 1 in Winter (1981).}. 
Pauli made such prediction in 1930 
in his famous letter in order to explain the continuous electron 
energy distribution in nuclear beta decay. 
It became immediately clear that neutrinos will be difficult to observe,
because the corresponding cross sections are so tiny. 
But in a series of experiments in 1953-59  Reines and Cowan (1953, 1959)
were able to prove convincingly that electron
anti-neutrinos from nuclear reactors are able to cause 
the inverse neutron beta decay, 
$\bar{\nu}_e + p \rightarrow e^+ +n$, and hence that they 
are real particles. Shortly afterwards, in 1962, 
the separate identity of the muon neutrinos,
$\nu_{\mu}$, was demonstrated (Danby  {\it et al.} 1962).
Another decade later, in 1975, the $\tau$ lepton was 
discovered (Perl {\it et al.} 1975) and the observation of its decay 
properties implied the existence of the third neutrino, $\nu_{\tau}$, that 
was directly observed only very recently (Kodama {\it et al.} 2001).
Precise measurements of the decay 
width of the $Z$ have shown that just three neutrino flavors 
($2.994 \pm 0.012$ from the combined fit to all LEP data (Groom {\it et al.} 2000))
participate in the weak interactions (at least for neutrinos with
masses less than $1/2 M_{\rm Z}$).

Phenomenologically, it is obvious that neutrinos of each flavor
are either massless or at least many orders of magnitude lighter 
than the corresponding
charged leptons with which they form the weak interaction doublets.
Based on these empirical facts,
the Standard Electroweak Model {\it postulates}
that all neutrinos are massless, and consequently have conserved
helicity (which is the same as chirality in this case) and that the
separate lepton numbers for electron, muon, and tau flavors are
conserved. Challenging this postulate of the vanishing neutrino mass
has recently become a central issue in many disciplines of
fundamental science, including particle and nuclear physics,
cosmology, and astrophysics. The present review is devoted
to one particular aspect of this broad effort. 

Ironically, while our knowledge of intrinsic neutrino properties
remains quite poor, these particles have been used as tools
to understand other phenomena. The tradition 
of underground neutrino detectors began thirty years ago when 
Davis and his collaborators were first able to detect neutrinos from
the Sun. (For the description of the history of solar neutrino
research see (Bahcall 1989).) This was, and 
together with the other experimental observations
of the solar neutrinos still is, the only clear proof that 
the basic energy generation in stars is understood. 
The birth of neutrino astronomy can be associated with the
observation of the neutrino burst from the supernova 1987A. 
Neutrino-induced reactions played an important role 
in establishing what is now known as the Standard Model
of electroweak interactions
when in 1973 the neutral currents were discovered via 
the observation of the $\nu_{\mu} + e \rightarrow  \nu_{\mu} + e$ 
scattering as well as the neutral current scattering 
of neutrinos on nucleons.
Finally, neutrinos have been extensively used in deep inelastic 
scattering experiments at CERN and FNAL, exploring the 
quark structure of nucleons.

The main problem in neutrino physics today is the question 
whether neutrinos, like all charged fermions, 
have a mass\footnote {For an up-to-date discussion of the neutrino
masses and the relevant experiments see (Fisher, Kayser, and
McFarland 1999) and (Zuber 1998).}.
Since direct kinematic tests of neutrino mass lack
at present the required sensitivity, the recent hints
for neutrino mass are indirect, based 
on the phenomenon of neutrino oscillations. If neutrinos are 
massive particles which behave in analogy 
to quarks, the states with a definite mass (i.e., the
``mass eigenstates'' which propagate as plane waves in vacuum) 
are not necessarily the partners of the charged leptons 
that couple to the vector bosons $W^{\pm}$ in doublets 
(i.e., the weak or flavor eigenstates)
\begin{equation}
\left( \begin{array}{c}
{\rm\nu_e} \\ {\rm e^-} \\ \end{array} \right) ~,
\left( \begin{array}{c}
\nu_{\mu} \\ \mu^- \\ \end{array} \right) ~,
\left( \begin{array}{c}
\nu_{\tau} \\ \tau^- \\ \end{array} \right) ~.
\end{equation}
The weak eigenstates $| \nu_l \rangle$ will be in such a case 
linear superpositions of the mass eigenstates $| \nu_i \rangle$
\begin{equation}
| \nu_l \rangle = \sum_i U_{l,i} | \nu_i \rangle ~,
\end{equation}
where the coefficients $U_{l,i}$ form the 
leptonic mixing matrix. If we assume that only three
neutrinos can contribute in the Eq.(2) above, 
then $U$ is a unitary $3 \times 3$ matrix\footnote{
Sometimes more than 3 mass eigenstates are considered.
The additional neutrinos must be sterile or heavy, i.e., they
must not participate in weak interactions so that the
constraint from the invisible width of the $Z$ boson 
is obeyed.}.

If Eq.(2) is valid, we encounter the phenomenon 
of neutrino oscillations in which a neutrino that was 
initially in the weak eigenstate $l$ can be spontaneously transformed, 
at least in part, into another weak eigenstate neutrino of flavor $l'$. 
(The idea of oscillations was discussed early on 
by Pontecorvo (1958, 1967)
and by Maki, Nakagawa and Sakata (1962).)

To see how that transformation happens,
recall that the mass eigenstate $| \nu_i  \rangle$ propagates 
according to the expression
\begin{equation}
| \nu_i  (t) \rangle   =   
e^{-i(E_i t - p_i L )} | \nu_i (0) \rangle              
 \simeq  e^{-i(m_i^2/2E) L} | \nu_i (0) \rangle ~,
\end{equation}
where $L$ is the flight path and in the last expression we assumed 
that the laboratory momenta and energies are
much larger than the neutrino rest masses $m_i$.
Let us consider now the propagation of a neutrino
which was created at $L=0$ as a weak eigenstate $| \nu_l  \rangle$.
At a distance $L$ this state is described by
\begin{eqnarray}
| \nu_l (L) \rangle & \simeq & 
\sum_i U_{l,i} e^{-i(m_i^2/2E)L} |\nu_i \rangle \nonumber \\
& \simeq & \sum_{l'} \sum_i U_{l,i} e^{-i(m_i^2/2E)L} U_{l',i}^* | 
\nu_{l'} \rangle ~.
\end{eqnarray}

In the last expression we used the inverse transformation to Eq.(2),
i.e., from the mass eigenstates back to the weak eigenstates. This
step must be taken since the only way one can detect neutrinos
is through their weak interactions. And in order to detect
the neutrino flavor we have to use the charged current weak
interactions, characterized by the production of the charged
leptons $| l' \rangle$.

Thus, the neutrino of flavor $l$ 
acquired components corresponding to other flavors $l'$. This is a purely
quantum mechanical effect, a consequence of the coherence in the
superposition of states in Eq.(2). The probability that the ``transition''
$l \rightarrow l'$ happens at $L$ is obviously
\begin{eqnarray}
& &P(\nu_l \rightarrow \nu_{l'})  = 
| \sum_i U_{l,i} U_{l',i}^* e^{-i(m_i^2/2E)L} |^2  \nonumber \\
& = &
\sum_i | U_{li} U_{l'i}^*|^2
+ \Re \sum_i \sum_{j \ne i} U_{li}U_{l'i}^*U_{lj}^*U_{l'j}
e^{i\frac{|m_i^2 - m_j^2|L}{2p}} ~.
\end{eqnarray}
This is an oscillating function of the distance $L$. 
The oscillation length depends
on the differences of the neutrino mass squares, $|m_i^2 - m_j^2|$,
and the oscillation amplitude depends on the mixing matrix $U$. 

The necessary and sufficient conditions for the existence
of neutrino oscillations is then the nonvanishing value of
at least one neutrino mass $m_i$ and the nonvanishing value
of at least one nondiagonal matrix element of the mixing matrix $U$.
If these conditions are fulfilled, the individual lepton 
flavor numbers
(electron, muon, tau) are no longer conserved. 

There is no fundamental theory which would allow us to deduce
the parameters describing the mixing matrix  $U$ and the
mass differences $\Delta m_{ij}^2$. These unknown parameters
must be determined empirically, by various neutrino
oscillation experiments. Such analysis is often performed 
in a simplified way by assuming
that only two neutrino flavors mix, e.g. $e$ and $\mu$. 
The mixing matrix $U$ then
depends only on one mixing angle $\theta$, and the oscillation probability,
Eq. (5), is also simplified
\begin{eqnarray}
U & = & \left( \begin{array}{rr} 
{\rm cos}\theta & {\rm sin}\theta \\  -{\rm sin}\theta & {\rm cos}\theta
\end{array} \right) ~,  \nonumber \\
P({\rm \nu_e} \rightarrow \nu_{\mu} , L) & = & 
{\rm sin}^22\theta {\rm sin}^2(\Delta m^2 L/4E) ~.
\label{eq:osceq}
\end{eqnarray}
Here $\Delta m^2 \equiv m_1^2 - m_2^2$, and we assume,
as before, that the neutrinos are ultrarelativistic. 
The probability that $\rm\nu_e$ remains
$\rm\nu_e$ is obviously 
\begin{equation}
P({\rm\nu_e \rightarrow \nu_e} , L) = 
1 - P({\rm\nu_e} \rightarrow \nu_{\mu} , L).
\label{eq:oscdiag}
\end{equation}

In this two-flavor scenario the oscillation amplitude is 
${\rm sin}^22\theta$ which vanishes if $\theta = 0$ or 90$^{\circ}$ and is
maximum if $\theta = $45$^{\circ}$. The oscillation length is
\begin{equation}
L_{osc} = 2\pi \frac{2E_{\nu}}{\Delta m^2} = 
\frac{2.48 E_{\nu} {\rm (MeV)}}{\Delta m^2{\rm (eV^2)}} {\rm meters} ~.
\label{eq:oscleq}
\end{equation}

To test for oscillations, one can perform either an {\it appearance} search in
which one looks for a new neutrino flavor (i.e.,
the deviations of  $P({\rm\nu_e} \rightarrow \nu_{\mu} , L)$
from zero), or a  {\it disappearance} 
test in which one looks for a change in the flux normalization
(i.e., the deviation of $P({\rm\nu_e \rightarrow \nu_e} , L)$ from unity). 
In either case, tests performed
at distance $L$ are only sensitive to the values of $\Delta m^2$ for which
$L \geq O(L_{osc})$. Or, in other words, neutrino
oscillations are observable only when $\Delta m^2 L/E \sim O(1)$.

So far we have considered only propagation of neutrinos in a vacuum. When
neutrinos propagate in matter, such as in the solar interior, the oscillation
pattern may be modified. This happens because electron neutrinos can
forward scatter on electrons by charged current interactions, and other
neutrino flavors cannot. Under favorable circumstances a resonance 
enhancement of the oscillation amplitude, the so-called 
Mikheyev-Smirnov-Wolfenstein (MSW) effect (Wolfenstein 1979, 1980)
and  (Mikheyev and Smirnov 1986a, 1986b), can take place.
Analogous matter induced oscillations can distinguish the
hypothetical sterile neutrinos, which have no weak interactions
at all, and the $\nu_{\mu}$ or $\nu_{\tau}$ neutrinos which
interact with matter (electrons and quarks) by the neutral current
weak interaction.
Neither of these kinds of matter effects is relevant for reactor
neutrinos.

For completeness, it is worthwhile to mention here two other issues
important in the study of the neutrino intrinsic properties.
One of them deals with the charge conjugation properties
of the neutrinos. Unlike the charged leptons, which are
Dirac particles, with distinct antiparticles, neutrinos can
be either Dirac or Majorana particles. In the latter case
of truly neutral neutrinos,
there is no distinction between the neutrinos and their
antiparticles, and even the total lepton number is not conserved.
In order to decide between these two possibilities, one has
to look for processes that violate the total lepton number,
such as the neutrinoless double beta decay. Other processes
of this kind, like the $\nu_e \rightarrow \bar{\nu}_e$ oscillations
(e.g. in the present context the emission of $\nu_e$ from
the nuclear reactor) are typically kinematically supressed, 
and their observation is  unlikely in forseeable future. 
The difference between the Dirac and Majorana neutrinos, while
of fundamental importance, does not influence the results
of the reactor oscillation searches described below.

The other issue worth mentioning is the possibility of the
$T$  or  $CP$ violation in neutrino oscillations (Cabibbo 1978
and Barger, Whisnant and  Phillips 1980).
In order to establish violation of
$T$ or $CP$ one would have to show that
\begin{equation}
P(\nu_{\ell'} \rightarrow \nu_{\ell}) 
\neq P(\bar{\nu}_{\ell'} \rightarrow \bar{\nu}_{\ell})~,
\end{equation}
i.e., that for example the probability of $\nu_{\mu}$ oscillating
into $\nu_e$ is different from the probability
of $\bar{\nu}_{\mu}$ oscillating into $\bar{\nu}_e$.

For the usual case of three neutrino flavors, one can
parametrize  the lepton mixing matrix
in terms of the three angles $\theta_1 = \theta_{13} $,
$\theta_2 = \theta_{23}$, and $\theta_3 = \theta_{12}$
and the $CP$ violating phase $\delta$.
\begin{eqnarray}
& & \left( \begin{array}{c}
\nu_e \\ \nu_{\mu} \\ \nu_{\tau}
\end{array} \right)
=  \\ 
& & \left( \begin{array}{ccc}
c_1c_3 & c_1s_3 & s_1e^{-i\delta} \\
-c_2s_3-s_1s_2c_3e^{i\delta} & c_2c_3-s_1s_2s_3e^{i\delta} &
c_1s_2 \\
s_2s_3-s_1c_2c_3e^{i\delta} & -s_2c_3-s_1c_2s_3e^{i\delta} &
c_1c_2
\end{array} \right)
\left( \begin{array}{c}
\nu_1 \\ \nu_2 \\ \nu_3
\end{array} \right)  \nonumber
\end{eqnarray}

where e.g. $c_1 = \cos\theta_1$ and $s_1 = \sin\theta_1$, etc.

The magnitude of the $T$ or $CP$ violation is characterized
by the differences
\begin{eqnarray}
& & P(\bar{\mu}\rightarrow\bar{e})-P(\mu \rightarrow e) =
-[P(\bar{\mu} \rightarrow \bar{\tau})-P(\mu \rightarrow \tau)] =
 \nonumber \\
& & P(e \rightarrow \tau)-P(\bar{e} \rightarrow \bar{\tau}) =
 \nonumber \\
& &-4c_1^2s_1c_2s_2c_3s_3\sin\delta[\sin\Delta_{12} + \sin\Delta_{23}
+  \sin\Delta_{31}] ~,
\end{eqnarray}
where $\Delta_{ij} = (m_i^2 - m_j^2) \times L /2E$.

Thus, the size of the effect is the same in all three channels,
and $CP$ violation is observable
only if all three masses are different (i.e., nondegenerate),
and all three angles are nonvanishing.
As will be shown below, reactor experiments constrain
the angle $\theta_1$ (or $\theta_{13}$),
to be small ($\sin^2 2\theta_{13} \le 0.1$). If that mixing
angle vanishes exactly, no $CP$ violation is observable
in the lepton sector,
independently of the value of the $CP$ violating phase $\delta$. 
To further improve the sensitivity to $\sin^2 2\theta_{13}$ is, therefore,
a matter of utmost importance.

\section{Physics motivation of modern experiments}

\begin{figure}[htb]
\epsfysize=6.5in \epsfbox{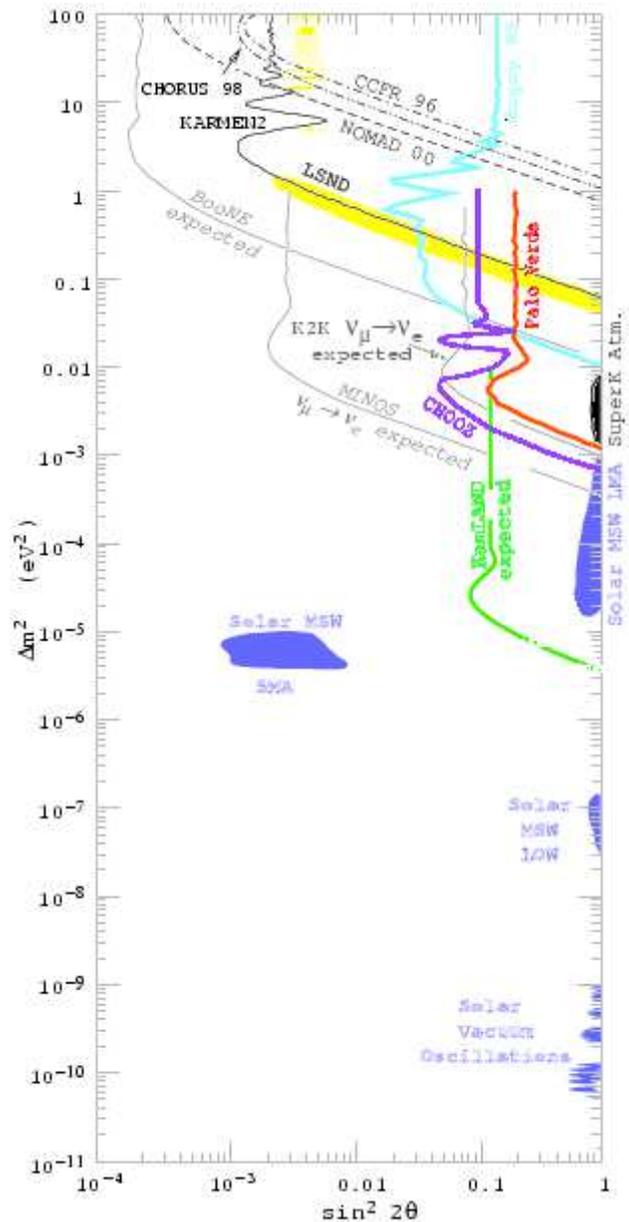}
\caption{ Phase-space for neutrino oscillations.
The existing limits on $\rm \nu_e - \nu_{\mu}$ are compared with current and
future experiments and the regions obtained by interpreting
the solar, atmospheric and
LSND neutrino anomalies as due to oscillations
(some of these effects
 are not necessarily   $\rm \nu_e - \nu_{\mu}$ oscillations.)
The MSW mechanism is used in plotting some of the solar neutrino regions.
The sensitivity of reactor experiments
is the same for $\nu_{\rm e} - \nu_{\tau}$
oscillations.   Limits are at 90\%~CL. }
\label{fig:sensi}
\end{figure}

Numerous searches for neutrino oscillations 
were performed during the last two decades using nuclear reactors as well
as particle accelerators as sources.
Since most of them have not observed evidence for neutrino
oscillations, their results are usually based on the 
simplified two neutrino mixing scenario and presented as 
an ``exclusion plot'', i.e., based on them certain
ranges of the parameters $\Delta m^2$ and sin$^22\theta$
can be excluded from further considerations as shown in 
Figure~\ref{fig:sensi}. 
However, at the present time there are three groups of measurements that
suggest the existence of neutrino oscillations. 
(And, at the same time, the parameter ranges suggested
by them are not excluded.) 
Only these {\it positive} results will be briefly discussed 
here, other experiments are listed in the 
Review of Particle Properties (Groom  {\it et al.} 2000).

\subsection{Experimental indications for neutrino oscillations}

The most prominent group of measurements which are commonly interpreted 
as evidence for neutrino oscillations
are often referred to as the ``atmospheric neutrino anomaly''
(Kajita and Totsuka 2001).
Primary cosmic rays impinging on the nitrogen and oxygen nuclei
at the top of the earth's atmosphere produce mostly pions,
which subsequently
decay via the chain 
$\pi^- \rightarrow \mu^- \bar{\nu}_{\mu}, 
\mu^- \rightarrow  e^- \bar{\nu}_e \nu_{\mu}$ (and the analogous
chain with $\pi^+$ etc.).
At sufficiently low energy, when such chains can fully develop,
the resulting atmospheric neutrinos therefore are expected to
follow the $\nu_{\mu} : \nu_e = 2 : 1$ ratio, which is essentially
independent of the details of the complicated process that created 
them. In addition, in an underground detector, one can deduce
the direction of the incoming neutrinos from the direction of the 
leptons ($e$ and $\mu$) created by the charged current interactions. 
Again, one is reasonably confident
that this zenith angle distribution can be accurately predicted.
If the $\nu_{\mu}$ and/or $\nu_e$ neutrinos oscillate, one expects
deviations from the 2:1 ratio mentioned above. Also, since the zenith
angle is simply related to the neutrino path length, one expects
deviations from the expected zenith angle dependence of the lepton
yield.

Both signatures of neutrino oscillations were in fact observed. The 
$\nu_{\mu}/\nu_e$ ratio is noticeably smaller, only about 60\%, than
the expected value. This result has been confirmed in four detectors
thus far.
The anomalous zenith angle dependence was first observed in
Kamiokande, and has been now confirmed, with much better statistical
significance, by SuperKamiokande 
(Kajita and Totsuka 2001, Fukuda 1998, Fukuda 2000).
If these effects indeed signify neutrino oscillations (and we do not 
know of
another viable explanation) then the corresponding mixing angle is large,
sin$^22\theta \approx 1$, and the value of the mass parameter 
$\Delta m^2$  is  in the range
10$^{-2}$ - 10$^{-3}$ eV$^2$. 
While the preferred scenario at present
involves $\nu_{\mu} \rightarrow \nu_{\tau}$ oscillations, it is not
clear that $\nu_{\mu} \rightarrow \nu_e$ oscillations are fully
excluded.

The second set of measurements that can be interpreted as  evidence
for neutrino oscillations
deals with the ``missing'' solar neutrinos
(Kirsten 2000, latest results in Fukuda 2001
and in particular in Ahmad 2001). The Sun produces
an intense flux of electron neutrinos as a byproduct
of the fusion reactions which generate solar power.
It is believed that the solar structure is understood
sufficiently well so that the flux and energy spectrum of the
neutrinos can be confidently predicted. The solar
neutrino fluxes have been measured in six experiments
so far. All of them report a deficit, i.e., the measured flux
is less than the expected one. Moreover, the reduction
depends on the neutrino energy, inferred experimentally from the
thresholds of the individual detectors.  The only viable 
explanation of the deficit appears to be neutrino
oscillation ($\nu_e$ disappearance).
The hypothesis that solar $\nu_e$ indeed oscillate into `active'
neutrinos that scatter on electrons via the neutral current weak
interaction is supported, at the 3 $\sigma$ level, by combining
the pure charged current measurements of SNO (Ahmad 2001) with
the charged + neutral current measurement of Super-Kamiokanke
(Fukuda 2001).

By contrast to
the attempts to explain the deficit by modification of
the solar model, which are unsuccessful, all existing
data can be simply and elegantly explained by invoking
neutrino mass.
In particular, the solution based on
the MSW effect
offers the most popular scenario.  Treating the problem
in the two-flavor framework explained above, one arrives
at several isolated islands in the $\Delta m^2$ -- 
sin$^22\theta$ plane. Two solutions correspond
to $\Delta m^2 \approx$ 10$^{-5}$ eV$^2$. One of them (
``small mixing angle'' or SMA)
has sin$^22\theta \approx $10$^{-2}$, while the other
one (``large mixing angle'' or LMA) has 
sin$^22\theta \geq $ 0.5. This latter solution, currently
giving the best fit to the data,  spans an
interval of $\Delta m^2$ extending up to 
10$^{-4}$ eV$^2$. The other possibilities have large mixing
angles and  $\Delta m^2 \approx$ 10$^{-7}$ eV$^2$ (LOW)
or  $\Delta m^2 \approx$ 10$^{-10}$ eV$^2$ (vacuum).

Finally, the only indication for oscillations involving man-made neutrinos
comes from the LSND experiment which finds evidence for the
$\bar{\nu}_{\mu} \rightarrow \bar{\nu}_{\rm e}$ and, with more limited
statistics, also for $\nu_{\mu} \rightarrow \nu_{\rm e}$
(Athanassopoulos {\it et al.} 1995, 1996, 1998). 
The former channel
uses neutrinos from the pion and muon decay at rest, with energies
less than $m_{\mu}/2$. The latter channel uses neutrinos from the pion
decay in flight which have somewhat higher energies. These are
appearance experiments; the observed signal should be absent if neutrinos
do not oscillate. The well determined quantity is the oscillation probability,
which has the value of about 3$\times 10^{-3}$. This result has not been
independently confirmed. An analogous experiment
which also uses the neutrinos from the pion and muon decay
at rest, \textsc{Karmen} 
(Armbruster {\it et al.} 1998,  Eitel 2000), found no evidence for the
$\bar{\nu}_{\mu} \rightarrow \bar{\nu}_{\rm e}$ oscillations.
However, the parameter space compatible with the LSND signal is
not fully excluded by  \textsc{Karmen}.

As we can see from this brief discussion, the last decade brought us a number
of clues. With the exception of the LSND signal, 
they all came from measurements
involving neutrinos produced by natural sources outside of our control.
A number of new experiments has been performed or are in various stages of 
planning in order to investigate further these tantalizing effects. 
Reactor experiments play an all-important role in this quest, owing to their 
unique ability to investigate very small neutrino-mass differences.

Like in many other aspects of neutrino physics 
there is a fundamental difference between 
the past reactor oscillation 
experiments\footnote{
For a general discussion of short-baseline
reactor experiments see Boehm and Vogel 1992 and Boehm 2001;
for the individual experiments see Kwon {\it et al.} 1981, 
Zacek  {\it et al.} 1986,  Achkar 1992, Achkar {\it et al.} 1995 and 1996,
 Vidyakin {\it et al.} 1994, Alfonin {\it et al.} 1998.}
and the more 
recent experiments with baselines of 1~km or more: 
experiments in this latter category are 
designed to further investigate, in a controlled environment 
with man-made neutrinos, 
particular regions of the oscillation parameter space 
where there are indications for
oscillations from other experiments.   
Hence the results from the new generation of reactor \nuebar 
detectors directly impact our understanding of the neutrino mixing matrix.

\subsection{Reactor versus accelerator-based oscillation experiments}

\begin{figure}[h]
\begin{center}
\epsfysize=3.3in \epsfbox{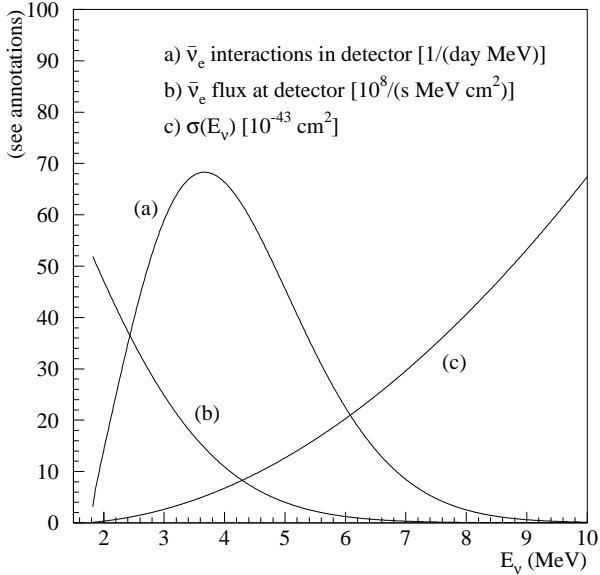}
\caption{ Reactor \nuebar flux, inverse beta decay cross section,
and \nuebar interaction spectrum at a detector based on such reaction.}
\label{fig:flux_and_xsec}
\end{center}
\end{figure}

Nuclear reactors produce isotropically \nuebar in the 
$\beta$ decay of the neutron-rich fission fragments.  
All detectors optimized for oscillation searches take advantage
of the relatively large cross-section and specific signature of the 
inverse-$\beta$-decay reaction 
$\bar{\nu}_e + p \rightarrow n + e^+$.   
Such cross-section is shown
in Figure~\ref{fig:flux_and_xsec} as function 
of the neutrino energy along with 
the neutrino flux at the reactor and 
the resulting interaction rate in a detector. We note
here that the detection reaction has a threshold of about
1.8~MeV.
Many of the merits and limitations of 
reactor-based experiments can be understood
observing that the energy of \nuebar is rather low, in the few-MeV range.
It directly follows that reactor-based experiments 
can only be of {\nuebar}-disappearance
type since the neutrino ``beam'' does not have 
sufficient energy to produce muons
(or taus) and the neutral-current reactions of 
the ``oscillated'' $\bar\nu_{\mu}$
or $\bar\nu_{\tau}$ have too low cross-section
and are un-distinguishable from
the many backgrounds present.  This first limitation 
makes {\rb} experiments well suited
only for investigating relatively large mixing angles.   
In practice experiments
have reported mixing sensitivities 
around 10\% at large $\Delta m^2$ (although the 
proposal for a very ambitious experiment 
with sensitivity better than 2\% at a 
particular $\Delta m^2$ will be discussed later).
The second limitation of {\rb} oscillation searches 
derives from the 

\begin{figure}[h]
\begin{center}
\epsfysize=3.5in \epsfbox{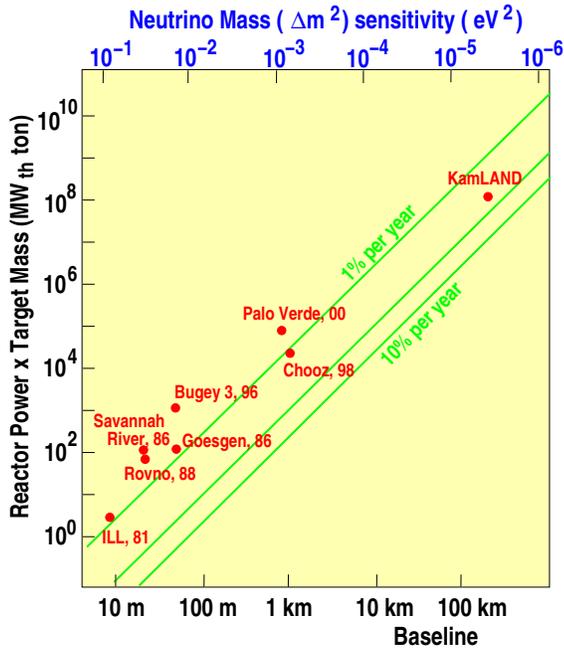}
\caption{ Neutrino $\Delta m^2$ sensitivity as a function of total
reactor power and detector fiducial mass
for detection based on the inverse-$\beta$ reaction
discussed in the text.   The baseline scales with
the $\Delta m^2$ sensitivity sought
according to Eq.~(\protect\ref{eq:oscleq}).
The fiducial-mass$\times$power necessary for the
experiment grows with the square of the baseline.
The past experiments are labelled by the name of the
reactor complex used. The approximate year of the experiment
is also indicated to show that the increased baseline and
$\Delta m^2$ sensivity followed more or less the chronological order.}
\label{fig:rate_mass}
\end{center}
\end{figure}

\noindent
fact that the only
known method of collimating neutrino beams employs 
the Lorentz boost of the parent 
particles from which decay the neutrinos are produced.  
For this reason low energy 
neutrinos are generally produced over large solid angles, 
while high energy ones may 
come in relatively narrow beams.   
Obviously a reactor emits  $\bar{\nu}_e$ in a
completely isotropic way, and this, 
together with the modest interaction cross-sections
available at low energy, makes the specific signal 
rates rather low.     
At the same time, however,
the low energy neutrinos provide us with 
a unique opportunity to probe the lowest regions
of $\Delta m^2$ that are otherwise beyond the reach of \ab searches.
Some of these tradeoffs are well illustrated by 
Figure~\ref{fig:rate_mass} where the $\Delta m^2$
sensitivity is shown, together with the necessary baseline, 
versus the reactor power and
detector fiducial mass for different statistical accuracies.

\indent
Oscillation searches using reactors as sources 
are particularly important today since
several of the indications for neutrino oscillations 
shown in Figure~\ref{fig:sensi} point
to regions of the parameter space at very small 
$\Delta m^2$ and nearly-full mixing.
Hence two \rb  experiments,
\textsc{Chooz} and \textsc{Palo Verde}, were performed to investigate 
the phenomenon of atmospheric 
neutrinos as $\bar{\nu}_e \rightarrow \bar{\nu}_x$ oscillations.  
Such experiments, described in 
detail below, had baselines of about 1~km 
and fiducial masses of the order of 10~tons.
For comparison, the much more complex
\ab \textsc{Minos} project between FNAL 
and the Soudan mine (Wojcicki 2001a)
and  analogous projects between CERN and Gran Sasso,
\textsc{Opera} and \textsc{Icarus} (for a brief description, see
e.g. Wojcicki 2001b),
will access similar $\Delta m^2$ values 
with GeV-energy neutrinos and a baseline of
the order of 1000~km.    However, the 5400 ton \textsc{Minos} detector
and its analog at Gran Sasso 
will be able to investigate also oscillation channels 
not including \nuebar and reach
a mixing parameter sensitivity substantially better than 1\%.

The \rb \textsc{Kamland} experiment, with a baseline 
larger than 100~km, will offer the unique 
opportunity of testing, with man-made neutrinos, 
the large-mixing-angle MSW solution of the 
solar neutrinos puzzle.   In this case 
the restriction to $\bar{\nu}_e \rightarrow \bar{\nu}_x$-oscillations
does not limit the interest of the experiment 
(since solar neutrinos do certainly involve
$\nu_e$ ), while its $\Delta m^2$ sensitivity is 
well beyond what can be practically achieved by accelerators 
(for comparison similar
$\Delta m^2$ sensitivity could be achieved in an \ab 
experiment with baselines of order
$10^5$~km, larger than the diameter of the earth).

Of course, the relatively lower energy of neutrinos 
from reactors pushes the optimization of 
{\rb} experiments to concentrate on the reduction 
and rejection of backgrounds from
natural radioactivity that is, on the other hand, 
hardly an issue in \ab detectors.
In this respect the correlated signature of 
the inverse-$\beta$ process,
the detection of the $e^+$ and neutron,  
plays a very important role.

While in the case of neutrinos produced by accelerators 
the experimenter has full control
over the status of the beam, the flux of \nuebar 
cannot be changed at will in commercial 
power nuclear reactors.   However, in practice, 
typical reactor optimization requires 
a refueling shutdown every 12 to 24
months.   Such shutdowns usually last about a month, 
providing a  convenient 
flux modulation that can be used to validate background subtraction methods.  
As explained in detail later, even in the 
case of \textsc{Kamland} that observes the neutrinos 
from about 70 reactor-cores, a substantial
flux modulation is provided by the coincidence of scheduled 
refueling outages in the 
Spring and Fall periods, when electricity demand is lowest.

Finally, we remark here that the fully isotropic flux 
produced by nuclear reactors 
eliminates the problems related with beam pointing that 
are present in experiments
using accelerators.   While the pointing accuracy 
required in these experiments is
well within the present technology, 
a fool-proof cross check of the beam-detector
alignment is certainly not trivial to obtain.

In conclusion \rb and \ab experiments offer 
complementary approaches to the quest for 
neutrino oscillations.  It is likely that only 
the combined~efforts on these two fronts,
together with other studies such 
as the search for neutrino-less-double-$\beta$ decay,
will allow us to elucidate the problem of mixing in the~lepton~sector.

\section{Reactor neutrino spectrum and flux determination}

\begin{figure}[h]
\begin{center}
\epsfysize=3.0in \epsfbox{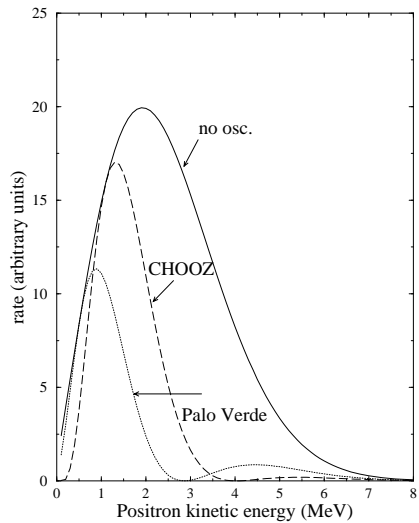}
\caption{ Expected positron energy spectra for
no oscillations (full line) and oscillations with parameters
$\Delta m^2 = 7.2\times 10^{-3}$~eV$^2$
and $\sin^2 2\theta = 1$ at the \textsc{Chooz} ($L\simeq 1$~km)
(dashed line)
and \textsc{Palo Verde} ($L\simeq 0.8$~km) (dotted line) experiments.
Adapted from Harrison, Perkins and Scott (1996).
}
\label{fig:perkins}
\end{center}
\end{figure}

Since \rb oscillation experiments are of the disappearance type,
the accurate determination of the \nuebar spectrum and
its absolute normalization
are essential ingredients of the measurements.   
We note here that
for oscillation parameters well within 
the experimental sensitivity the evidence
for oscillations would manifest itself as 
a deficit of events accompanied by a distortion
of the energy spectrum as shown by 
the example
in Figure~\ref{fig:perkins}.
However, as the true value of 
the oscillation parameters moves closer to the sensitivity
boundary of the experiment, 
the spectral shape loses power and the accuracy of the measurements
essentially relies on the total event count and, 
hence, the knowledge of the absolute reactor flux.
This last scenario also corresponds to 
the more usual case in which no oscillations are 
observed and an upper-limit is set,
as well as to the case of large $\Delta m^2$ where
the spectrum distortions are washed out.

While in this section we will concern ourselves 
mainly with a-priori reactor \nuebar yield 
determinations, multiple-baseline measurements 
are possible and have been performed in the past
at the Goesgen (Zacek {\it et al.} 1986) and Bugey
(Achkar  {\it et al.} 1995, 1996) reactors.    
Indeed such measurements 
helped gaining confidence in the reactor yield estimates, and, 
although were not the main goal of
\textsc{Chooz} and \textsc{Palo Verde}, 
have been recently proposed (Mikaelyan 2000) 
for a more accurate determination of
the mixing angle $\theta_{13}$ for the atmospheric neutrino region.
\textsc{Chooz} could take advantage of the $\approx 115$~m distance
between the reactors for deriving weaker exclusion limits, that were however
less affected by systematics.  

\subsection{Anti-neutrino production}

The determination of the \nuebar yield proceeds, schematically, 
in three steps.    First, the 
thermal power of each reactor core is measured, 
accurately and essentially continuously. Based on
such measurements, and starting from
the initial fuel composition, the burn-up state 
can be computed as function of time.    
Small corrections due to other reactor parameters 
that modify the criticality of the 
core are also introduced at this time.   Reactor simulation 
codes are often used at this
stage and produce an accurate instantaneous 
fission rate for each of the relevant
isotopes through the fuel cycle.
In the second step the neutrino spectrum is derived from 
the fission rate.  
Finally, as the last step, 
the neutrino spectrum emitted by the reactors must be converted
into an estimate of the experimental observable, the positron spectrum in
the detector. Each of these steps will be explained in a 
separate subsection.
 
Typical modern 
commercial Light Water Reactors (LWR) have 
thermal powers of the order of 3~GW$_{th}$.   This figure 
applies to both Pressurized  
Water Reactors (PWR) and to the less common 
Boiling Water Reactor (BWR) designs.   In both
cases the fuel is enriched to 2-5\% in $^{235}$U.
Since on average each fission produces $\sim 200$~MeV 
and $\sim$~6~\nuebar we conclude that the
typical yield is $\sim 6\times 10^{20}$~{\nuebar}~core$^{-1}$~s$^{-1}$ 
(of course part of this
flux will be below the detection threshold, 
see Figure~\ref{fig:flux_and_xsec}).

It is easy to understand why  $\sim$~6~\nuebar are
produced per fission.
Take, as an example, the most common $^{235}$U fission, which
produces two unequal fragments, and typically two new neutrons
that sustain the chain reaction,
\begin{equation}
^{235}U + n \rightarrow X_1 + X_2 + 2n~.
\end{equation}

\begin{figure}[h]
\begin{center}
\epsfysize=2.5in \epsfbox{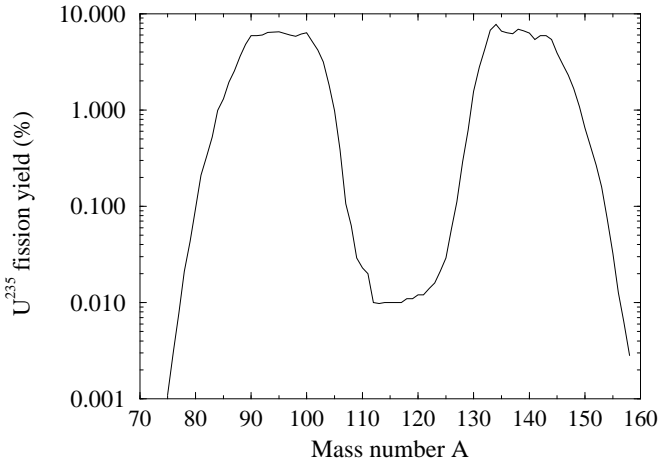}
\caption{ Yields (in \%) for $^{235}$U
thermal neutron fission (normalized to 2 for the two
fragments)}
\label{fig:u235f}
\end{center}
\end{figure}

The mass distribution of the fragments (so-called fission yields)
is shown in Fig. \ref{fig:u235f}. The lighter fragments have,
on average, $A \simeq 94$ and the heavier ones $A \simeq 140$.
The stable nucleus with $A = 94$ is $_{40}{\rm Zr}^{94}$ and the
stable $A = 140$ nucleus is $_{58}{\rm Ce}^{140}$. These two nuclei
have together 98 protons and 136 neutrons, while the initial
fragments, as seen from the equation above, have 92 protons
and 142 neutrons. To reach stability, therefore, on average
6 neutrons bound in the fragments have to $\beta$ decay, emitting
the required 6 $ \bar{\nu}_e$.

While the total number of $ \bar{\nu}_e$ is easy to estimate,
and can be accurately determined given the known fission
yields, their energy spectrum, which is of primary interest
for the oscillation searches discussed here, requires more
care. In particular, the commonly used
neutrino detection reaction, the inverse neutron
$\beta$ decay, has $\sim$1.8~MeV threshold.
Only about 1.5 $ \bar{\nu}_e$/fission (i.e. $\sim$ 25\%)
of the total are above that threshold and hence can be
detected, hence the total of $\sim$6 \nuebar per fission
is irrelevant.

The existence of the 1.8~MeV threshold in the detection process 
$\bar\nu_e + p \rightarrow n + e^+$ 
automatically insures that 
 only \nuebar from large 
$Q$-value, and hence short half-life, $\beta$-decays are detected.
Thus the observed \nuebar signal tracks
closely in time the power excursions in the reactor.   
This is of some 
practical importance as large quantities of spent fuel 
are usually stored on-site by 
reactor operators.   There is no need to track the inventory 
of spent fuel and to worry about the  $\beta$-decays
of the neutron-activated reactor materials which
have typically a low $Q$-value and therefore long half-life products.
In practice, 
after a few hours from reactor turn on/off, the detectable \nuebar 
flux can be considered saturated.

\subsection{Fission-rates determination}

\begin{figure}[h]
\begin{center}
\epsfysize=3.0in \epsfbox{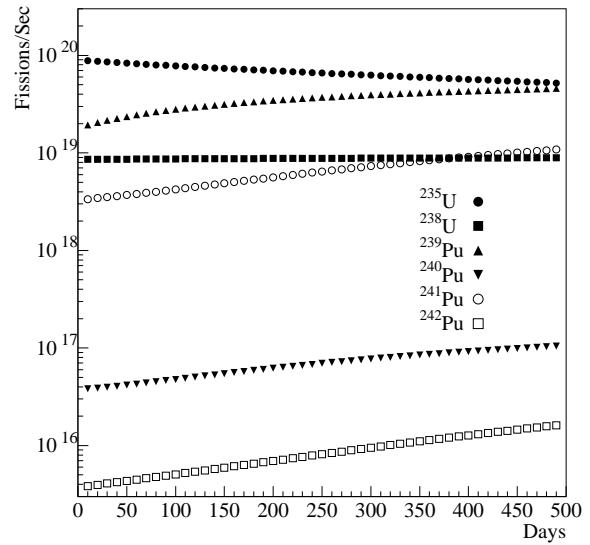}
\caption{ Time evolution of fission rates for each of the six
most important isotopes in one of the Palo Verde reactor cores.
The horizontal
scale covers a full fuel cycle, at the end of which about 1/3 of the core is
replaced with fresh fuel.   Only the four most important isotopes are normally
used to predict \nuebar yields.}
\label{fig:isotope_evolution}
\end{center}
\end{figure}

The four isotopes whose fission  is the source of virtually
all the reactor power are $^{235}$U, $^{238}$U, $^{239}$Pu, 
and $^{241}$Pu.   The fission rates deriving from their evolution 
during a typical fuel cycle in one of the Palo Verde reactors is shown in 
Figure~\ref{fig:isotope_evolution} as calculated by a core simulation 
program (Miller 2000).
For comparison we also show the evolution of $^{240}$Pu and $^{242}$Pu
that give the next to leading contributions.  The contribution of these
isotopes is of order 0.1\% or less and will not be considered further.

Each isotope produces a unique neutrino spectrum through 
the decay of its fission fragments and their daughters, so 
plutonium breeding results in 
a small but noticeable change in the emitted neutrino spectrum.

Two types of uncertainties can be attributed to 
the isotope compositions described in 
Figure~\ref{fig:isotope_evolution}: errors deriving from uncertainties in the 
initial fuel composition and in the measurement 
of the plant parameters that are
used as input to the simulation, 
and errors due to imperfect core and neutronics
modeling by the simulation program itself.
The errors intrinsic to the simulation 
are known to contribute by substantially 
less than 1\% to the neutrino yield from tests 
in which fuel is sampled and analyzed
for isotopic composition at the end of a fuel cycle.   

\begin{figure}[h]
\begin{center}
\epsfysize=3.0in \epsfbox{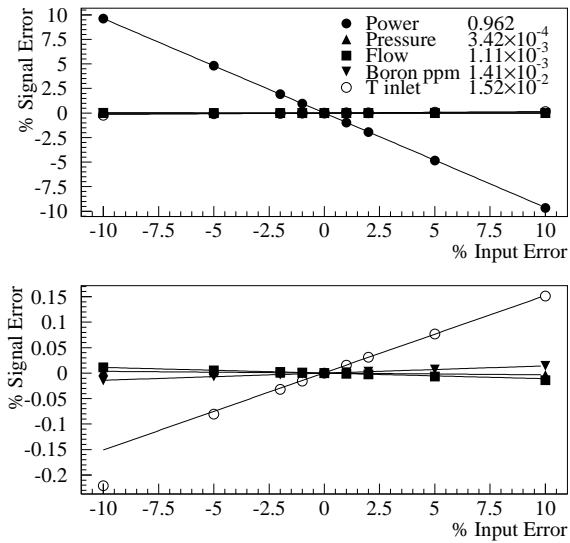}
\caption{ Correlation between \nuebar yield and the five most
important inputs to the core simulation for a PWR.
The numbers in the key are the slopes of the fitted lines. Note that
a variation of even 10\% in any of the parameters,
but power, has little effect
on the output of the simulation.}
\label{fig:rea_para}
\end{center}
\end{figure}

The correlation between the \nuebar yield 
and the plant parameters used as input
to the simulation is shown in Figure~\ref{fig:rea_para}.
Apart from the obvious correlation with the thermal power, 
other parameters enter
the simulation because they affect the criticality by altering the neutron
transport in the core (generally by the water density 
and boron absorber concentration).
We see that for the parameter 
with largest correlation besides power, the 
water temperature in the cold legs, an error of 10\% produce 
an uncertainty of only 0.15\% in the \nuebar
yield.   Of course the inlet temperature is known to much better that 10\%.

Economic and safety reasons provide plant operators 
with an incentive to accurately 
measure the thermal power of the reactors.    
Indeed, usually more than one method is
used and the results are compared to understand 
the size of the uncertainties.  
Calorimetric methods (simultaneous measurement 
of temperature and flow-rate of the water
outlet in the secondary cooling loop) 
give the smallest error ($\sim$ 0.6 - 0.7\% at 
\textsc{Chooz} and \textsc{Palo Verde}) 
and are used as primary power estimate.

\subsection{From fission rates to the \nuebar spectra}

The instantaneous fission rates of the four isotopes $^{235}$U, $^{238}$U, 
$^{239}$Pu, and $^{241}$Pu found above are then 
used as an input for the evaluation of the \nuebar spectrum.
For all but $^{238}$U  careful
measurements of the $\beta$ spectrum from fission 
by thermal neutrons were performed 
(Schreckenbach {\it et al.} 1985, Hahn {\it et al.} 1989). 
These are converted to neutrino spectra 
as explained below.  
However, Refs. (Schreckenbach {\it et al.} 1985, Hahn {\it et al.} 1989)  
do not include $^{238}$U 
that undergoes only  fast neutron fission 
and hence was not accessible to such measurements.

There are, at present, several methods available to evaluate
the $ \bar{\nu}_e$ spectrum.
For the $\bar{\nu}_e$ associated with $^{235}$U, $^{239}$Pu,
and $^{241}$Pu,
which undergo thermal neutron fission,
it is customary to use a hybrid method,
based on the conversion of the  measured
electron spectra
associated with the thermal neutron induced fission,
into the $\bar{\nu}_e$ spectra.
Clearly, the electron and $\bar{\nu}_e$ originate from the same
$\beta$ decay and  share the available endpoint
energy. For a single branch the conversion is therefore trivial.
However, in general there are many branches, and many nuclei
with different charges.
For electron and antineutrino energies well above the electron mass
the two spectra are quite similar (Schreckenbach {\it et al.} 1985)
\begin{equation}
\frac{{\rm d}N}{{\rm d}E_e} \simeq \frac{{\rm d}N}{{\rm d}E_{\bar{\nu}_e}} ~,
\end{equation}
where $E_e = \sqrt{p_e^2 + m_e^2}$ is the full electron energy,
and the proportionality constant deviates from unity by at most
5\%. Naturally, one would like to convert the spectra more
accurately.

Formally, the conversion
can be performed exactly as follows. Let $n(E,Z)$, assumed
to be a continuous function, describe the distribution
of endpoints $E$ and nuclear charges $Z$. The electron spectrum is then
\begin{equation}
 Y(E_e)  =  \int_{E_e}^{\infty} {\rm d}E
n(E,Z) k(E,Z) p_e E_e (E - E_e)^2 F(E_e,Z),
\label{eq:specint}
\end{equation}
where $ k(E,Z)$ is the spectrum  normalization constant,
$p_e$ is the electron momentum, and $F(E_e,Z)$
is the Fermi function describing the Coulomb effect on the
emitted electron.
Provided that the electron spectrum $Y(E_e)$ is 
measured, the endpoint
distribution can be determined,
\begin{equation}
n(E,Z) = - \frac{1}{2 k(E,Z)} \frac{{\rm d^3}}{{\rm d}E^3}
\left( \frac{Y(E)}{p E F(E,Z)} \right) ~.
\end{equation}
Once the distribution $n(E,Z)$ is known, the $\bar{\nu}_e$
spectrum is readily calculated by the integral analogous to
the Eq. (\ref{eq:specint}).
In (Davis  {\it et al.} 1979, Vogel  {\it et al.} 1981)
 it was shown that such conversion procedure
depends only very weakly  on the  value of $Z$.
In practice, an empirical relation $\bar{Z}(E)$ between the
average $\bar{Z}$ and the electron energy
has been used (Schreckenbach {\it et al.} 1985)
($\bar{Z} = 49.5 - 0.7E - 0.09E^2$ with $E$ in MeV).

When using the measured electron spectra the above expressions,
involving third derivatives, are obviously impractical. Instead,
the integral in Eq. (\ref{eq:specint})
is replaced by a finite sum of 30 hypothetical
beta decay branches with branching ratios $b_i$ and 
equidistant endpoints $E_0^i$,
\begin{equation}
Y(E_e) =  \sum_i
b_i k(E_0^i,\bar{Z}) \delta(E_e,E_0^i)
p_e E_e (E_0^i - E_e)^2 F(E_e,\bar{Z}),
\end{equation}
where $ \delta(E_e,E_0^i)$ describes the small outer radiative corrections.
One can now begin with the largest value of $E_0^i$ (only one branch)
and determine the corresponding branching ratio $b_i$
using the electrons of energies between that $E_0^i$
and the next smaller one, and continue
in this fashion step by step until the smallest $E_0^i$  is reached.
Possible variations in the number and distribution of the endpoints
 $E_0^i$ affects the resulting $\bar{\nu}_e$ spectrum not more than
at 1\% level (see Schreckenbach {\it et al.} 1985).
Having determined the set  $b_i, E_0^i $ it is trivial
to obtain
\begin{eqnarray}
Y(E_{\bar{\nu}_e}) & = &  \sum_i
b_i k(E_0^i,\bar{Z})
E_{\bar{\nu}_e}^2 (E_0 - E_{\bar{\nu}_e}) \nonumber \\
& \times & [(E_0 - E_{\bar{\nu}_e})^2 - m_e^2]^{1/2}
F(E_0 - E_{\bar{\nu}_e}, \bar{Z}) ~,
\end{eqnarray}
where the irrelevant radiative corrections were omitted. This is the
procedure used in deriving the neutrino spectra
associated with fission of  $^{235}$U, $^{239}$Pu,
and $^{241}$Pu which account for about 90\% of the reactor
$\bar{\nu}_e$.

The \nuebar spectra in 
Refs. (Schreckenbach {\it et al.} 1985, Hahn  {\it et al.} 1989,
Davis  {\it et al.} 1979, Vogel  {\it et al.} 1981,
Klapdor and Metzinger  1982a, 1982b)
are given as tables. A somewhat less accurate, but easier to implement,
analytical approximation is given in Vogel and Engel (1989).
(The fit error to the total rate is about 1.2\% for $^{235}$U
and only about 0.3\% for $^{239}$Pu and $^{241}$Pu (Miller 2000).)

For the \nuebar associated with the $^{238}$U fission
one has to use the 
straightforward  summation of the
spectra of the $ \bar{\nu}_e $ from all individual $\beta^-$
decays. Thus
\begin{equation}
\frac{{\rm d}N}{{\rm d}E_{\bar{\nu}}} =
\sum_n Y_n(Z,A,t) \sum_i b_{n,i}(E_0^i)P_{\bar{\nu}}(E_{\bar{\nu}},
E_0^i,Z) ~,
\label{eq:sumnu}
\end{equation}
where  $ Y_n(Z,A,t)$ is the number of $\beta$ decays per unit
time of the fragment $Z,A$ after the fissioning material has
been exposed to neutrons for a time $t$ and the label
$n$ characterizes each fragment. For $t$ larger than
the $\beta$ decay lifetime of the fragment $Z,A$ the quantity $Y_n$
converges toward the cumulative fission yield and becomes
independent of $t$. Naturally,  each fission fuel is characterized
by a different set of yields $Y_n$.

The quantities $ b_{n,i}(E_0^i)$ in Eq. (\ref{eq:sumnu}) are the
branching ratios for the $i$th branch with the maximal
electron energy (endpoint energy) $E_0^i$. The branching ratios
are normalized to unity.

Finally, the function $P_{\bar{\nu}}(E_{\bar{\nu}},E_0^i,Z)$ is the
normalized spectrum shape. It is usually assumed, here and in the
conversion method explained above, that all
relevant $\beta$ decays have the allowed shape,
\begin{eqnarray}
& & P_{\bar{\nu}}(E_{\bar{\nu}},E_0,Z)  =  k(E_0,Z) E_{\bar{\nu}}^2
(E_0 - E_{\bar{\nu}}) \nonumber \\
& \times & [(E_0 - E_{\bar{\nu}})^2 - m_e^2]^{1/2}
F(E_0 - E_{\bar{\nu}},Z).
\label{eq:ashape}
\end{eqnarray}
To use Eq. (\ref{eq:ashape}) is a very good
approximation in practice and causes a totally negligible error.

The weakness of this method is the incomplete information on
the endpoint distribution and branching ratios of some
fission fragments, in particular those with very short
lifetimes and high decay energies. These `unknown' decays
contribute as much as 25\% of the $\bar{\nu}_e$ at energies
above 4 MeV. In practice, nuclear models are used to
supplement the missing data. Examples of calculations based
on this method are 
(Davis  {\it et al.} 1979, Vogel  {\it et al.} 1981,
Klapdor and Metzinger  1982a, 1982b, Tengblad {\it et al.} 1989).
An example
of an extension to lower $\bar{\nu}_e$ energies,
where the neutron activation of the reactor materials
plays a role, is given by Kopeikin, Mikaelyan, and Sinev (1997).

While the summation method played an important role in the
early oscillation searches, at present it is needed only
for the description of the $\bar{\nu}_e$ from $^{238}$U
fission as pointed out above. 
That component contributes only about 11\% to
the neutrino signal. We show below that the error associated
with the summation method is less than 10\% and hence it
contributes less than 1\% to the overall uncertainty.

The ultimate check of the accuracy of the prediction outlined above consists
in comparing the results in terms of \nuebar energy spectrum with the
measurements performed in short baseline reactor oscillation experiments.

\subsection{From  \nuebar to positrons}

Since in all reactor experiments one measures the positron spectra,
and not directly the \nuebar spectra, one has to understand
quantitatively how these  are related.
In other words, one has to know the cross section of the
`detector' reaction \nuebar $+p \rightarrow e^+ + n$.

The total cross section for this reaction, neglecting terms of
order $E_{\nu}/M$, is given by the standard formula
\begin{eqnarray}
\sigma^{(0)}_{tot}  &  = &
\sigma_0\; (f^2 + 3 g^2)\; E_e^{(0)} p_e^{(0)}  \nonumber \\
&  = & 0.0952 \left(\frac{ E_e^{(0)} p_e^{(0)}}{1 {\rm\ MeV}^2}\right)
\times 10^{-42} {\rm\ cm}^2\,,
\label{eq:sigtot0}
\end{eqnarray}
where $ E_e^{(0)} = E_{\nu} - (M_n - M_p) $ is the positron energy
when the (small) neutron recoil is neglected, and
$p_e^{(0)}$ is the corresponding momentum. 
The vector and axial-vector coupling constants are
$f = 1, g = 1.26$ and
\begin{equation}
\sigma_0 = \frac{G_F^2 \cos^2 \theta_C}{\pi} (1 + \Delta_{inner}^R) ~,
\end{equation}
where the energy independent inner radiative corrections are
$ \Delta_{inner}^R \simeq 0.024$.

The cross section
can be expressed in terms of the
neutron lifetime and the phase space factor $f^R_{p.s.} =
1.7152$ (Wilkinson 1982) as
\begin{equation}
\sigma^{(0)}_{tot} =
\frac{2 \pi^2/m_e^5}{f^R_{p.s.} \tau_n}\; E_e^{(0)} p_e^{(0)}\,.
\end{equation}
In this way, the cross section is tied directly to the neutron
lifetime, known to 0.2\% (Groom {\it et al.} 2000), 
no knowledge of $G_F$, $f/g$ or the Cabibbo angle $\theta_C$
is in fact needed.

The (small) energy-dependent outer radiative corrections to
$\sigma_{tot}$ are given in Vogel (1984) and Fayans (1985).
The corrections to the cross section  of
order $E_{\nu}/M$, which are not negligible even for
the reactor energies, and the angular distribution of the
positrons  are described in Vogel and Beacom (1999).
The exact threshold of the reaction is
\begin{equation}
E_{\nu}^{thr} = \frac{(M_n + m_e)^2 - M_p^2}{2M_p} = 1.806~ {\rm MeV}
\end{equation}
instead of just $M_n + m_e - M_p$ = 1.804 MeV when the recoil is neglected.

\begin{figure}[h]
\begin{center}
\epsfxsize=3.4in \epsfbox{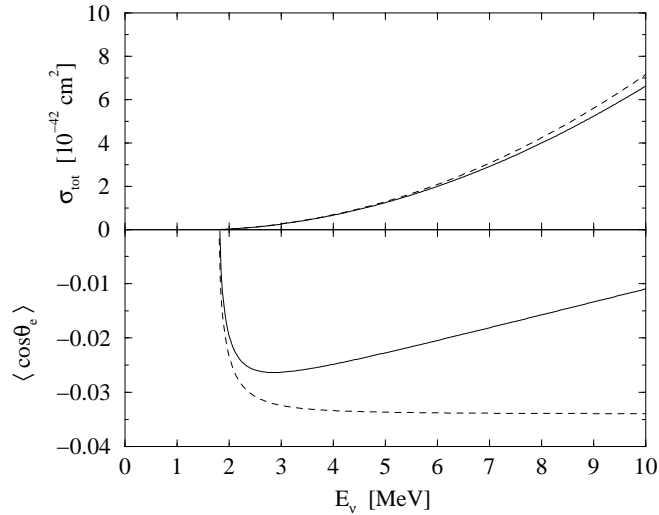}
\caption{Upper panel: total cross section for $\bar{\nu}_e + p
\rightarrow e^+ + n$; bottom panel: $\langle \cos\theta \rangle$;
as a function of the antineutrino energy.  The
solid line is the ${\cal O}(1/M_n)$ result and the short-dashed line is
the ${\cal O}(1)$ result, Eq.~\ref{eq:sigtot0}.}
\label{fig:reaction}
\end{center}
\end{figure}

Using the results of Vogel and Beacom (1999)
one can evaluate the total cross section
as well as the quantity $\langle \cos\theta_e \rangle$ which characterizes
the positron angular distribution 
essentially exactly. These quantities are shown in 
Figure \ref{fig:reaction}.
The high energy extension of the total and differential
cross section has been discussed already
in the classic paper by Llewellyn-Smith (1972). Near threshold, however,
that treatment must be modified as shown in Vogel and Beacom (1999).

The positron angular distribution, characterized by 
$\langle \cos\theta_e \rangle$ is rarely accessible.
It is of interest, however, to consider also the
angular distribution of the recoil neutrons that
are also detected.
Since in the laboratory system the proton is at rest,
the neutron is initially emitted at a forward angle
restricted by
\begin{equation}
\cos(\theta_n)_{max} = \frac{\sqrt{2 E_{\nu}\Delta - 
(\Delta^2 - m_e^2)}}{ E_{\nu}} ~,
\end{equation}
where $\Delta = M_n - M_p \sim $ 1.3 MeV. The average
$\langle \cos(\theta_n) \rangle$ 
is considerably closer to unity (Vogel and Beacom 1999).

It is often possible to localize the points
where the positron was created and where the neutron was
captured and
even though the neutron undergoes many elastic scatterings
before capture, its final position maintains some memory of its
original direction. Simulations suggest that the typical displacement
of the two vertices is $\langle x \rangle \sim$ 1.5 cm
in the organic scintillator.
In fact, in previous  reactor experiments
(Zacek {\it et al.}, 1986, Achkar  1992, Zacek 1984,  Achkar {\it et al.} 1995)
the neutron displacement was clearly observed, in the Goesgen
experiment in particular, at $\simeq 10 \sigma$
level. The same effect was also observed at  \textsc{Palo Verde}.
Moreover, the single vessel \textsc{Chooz}
experiment was able to measure the average neutron-positron separation
and to base on it a determination of the \nuebar incoming direction
with an uncertainty of $\simeq 8^{\circ}$ (Apollonio  {\it et al.} 2000).

Given a reliable simulation of the neutron transport, this asymmetry
allows, albeit with large errors, a direct measurement of the
detector background.
In case of \nuebar detection from a future Supernova, this
technique may provide, as shown by Apollonio  {\it et al.} (2000),
a crude 
but useful determination of the direction
to the \nuebar source, i.e. of the star position.

\subsection{Accuracy of the flux and spectrum predictions}

Once the cross section and $\bar{\nu}_e$ spectra are known,
the corresponding positron yield is easily evaluated. 
In reactor experiments, the neutron recoil is quite small
(10-50 keV) and thus the positron energy is simply
related to the incoming $\bar{\nu}_e$ energy,
\begin{equation}
E_{\nu} \simeq (E_e + \Delta) \left[ 1 + \frac{E_e}{M_p} \right]
+ \frac{\Delta^2 - m_e^2}{M_p} ~,
\end{equation}
where, as before, $ \Delta = M_n - M_p$ and we used $\cos\theta_e = 0$
as a good approximation of the average $\langle \cos\theta_e \rangle$.
We note here that possible detector efficiency dependence on the
positron energy requires special care.

\begin{figure}[h]
\begin{center}
\epsfxsize=2.8in \epsfbox{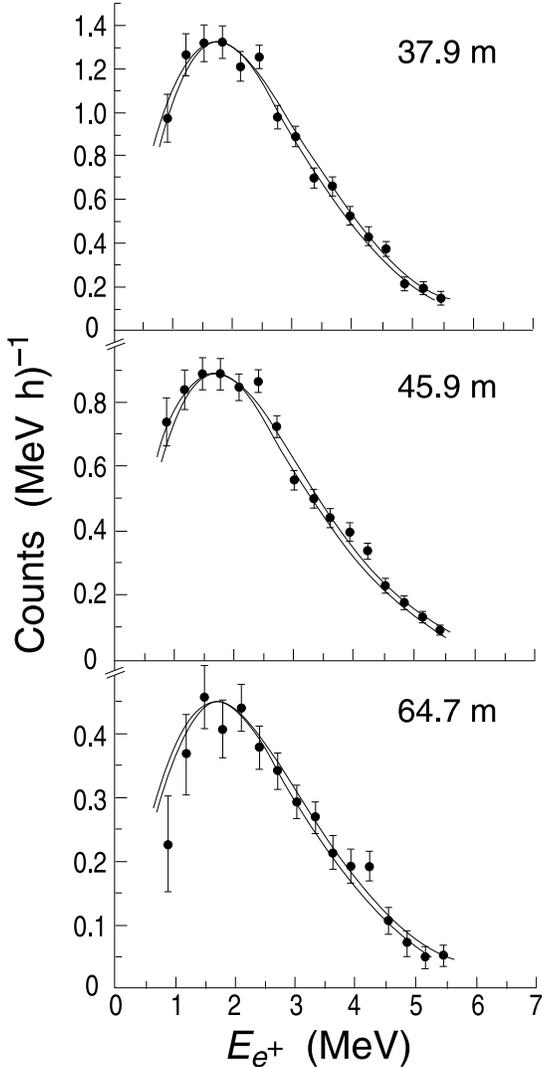}
\caption{ Positron spectrum
observed by the Goesgen experiment for
three different baselines (Zacek 1984). The two continuous lines
represent fits to the data and the predictions
obtained, as described in the text, using the measurements
(Schreckenbach {\it et al.} 1985, Hahn  {\it et al.} 1989)
for $^{235}$U, $^{239}$Pu, and  $^{241}$Pu
and theoretical calculations for $^{238}$U.}
\label{fig:goesgen}
\end{center}
\end{figure}

In order to check the accuracy of the prediction one has to
compare the results, in terms of \nuebar energy spectrum
and total flux normalization, with the 
measurements performed in short baseline reactor oscillation experiments.
Since such experiments have not reported the observation of oscillations, 
we can assume that their measurements represent the direct 
determination of the reactor
spectrum at production. 
There are four factors needed for the
evaluation of the expected positron yield and spectrum:
the distance to each core, the number of target protons,
the cross section for 
$\bar{\nu}_e + p \rightarrow n+e^+$, and of course the quantity 
one wants to test, the \nuebar spectrum at the source.

The distance to the reactors is trivially obtained with negligible 
uncertainty. The number of protons in the target requires knowledge of 
the chemical composition and mass of both scintillator and possible other
detector materials where \nuebar can be 
captured and recorded with finite efficiency.
Typically errors smaller than 1\% are achievable for this parameter.
The cross section has been discussed at the previous subsection,
and its uncertainty is also less than 1\%. Finally, the \nuebar
spectrum is the object of the test.

The \nuebar flux
normalization alone was tested in Declais {\it et al.} (1994) 
where the total neutron yield was measured with high
statistical accuracy, and  found to be in
agreement with the expectation based on the 
known neutron lifetime at the 3\% level.

The validity of the tests performed
at short distances from a reactor is reinforced by the fact that some
of the  experiments
such as Goesgen
(Zacek  {\it et al.} 1986) or Bugey
(Achkar  {\it et al.} 1995, 1996) did
measurements at different baselines observing 
no difference between the spectra.
This is shown for Goesgen in Figure~\ref{fig:goesgen}.   
Moreover, the
relatively recent Bugey~3 measurements were performed at 15-40~m distance from 
the core and recorded very high statistics 
(some $1.5\times 10^5$ \nuebar events).

\begin{figure}[h]
\begin{center}
\epsfxsize=4.0in \epsfbox{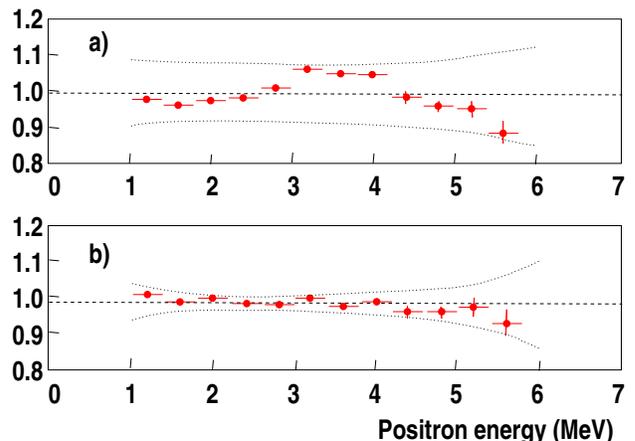}
\caption{ Ratio between Bugey~3
measurements and different predictions.
In a) the measurements are compared
to the a-priori calculations of Klapdor and Metzinger (1982a, 1982b).
In b) Bugey~3 data is compared to the prediction
obtained using the $\beta$ spectra measurements
of Schreckenbach {\it et al.} (1985) and Hahn  {\it et al.} (1989),
and the calculation mentioned for $^{238}$U.
The dashed envelopes are estimates of the overall systematics.
Adapted from Achkar  {\it et al.}  (1996).}
\label{fig:bugeyIII-theory}
\end{center}
\end{figure}

The good agreement between Bugey~3 data and the non-oscillation predictions 
is demonstrated in Figure~\ref{fig:bugeyIII-theory}.   
In panel a) the prediction is generated
purely from theory.   Of more practical importance 
is panel b) where the prediction derives
from $\beta$ spectra (except for $^{238}$U where theory is used).  In this case
a fit to a horizontal line gives a level 
of 0.99 with $\chi^2 / d.o.f. = 9.2/11$.     
The final Bugey~3 result is quoted as having a 1.4\% total uncertainty.

\begin{figure}[h]
\begin{center}
\epsfxsize=3.0in \epsfbox{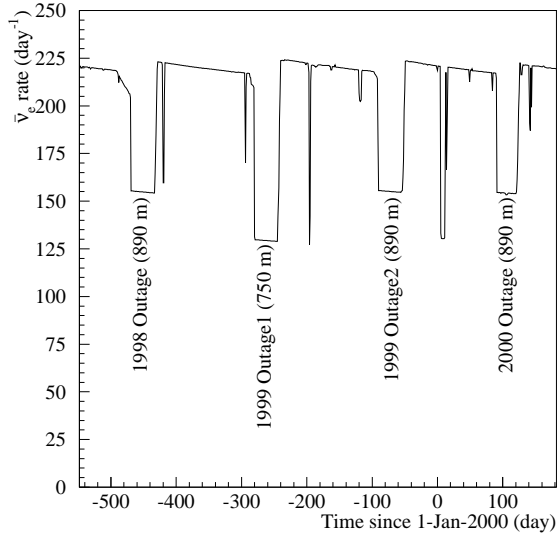}
\caption{Expected number of \nuebar
interactions in the \textsc{Palo Verde}
detector during the $\sim$2 years of data-taking of the experiment.
Note that one reactor
is closer to the experiment while the other two are equidistant; this
explains the
different excursion for one of the refuelings.    The steady decline in \nuebar
interactions during the cycle is the effect of fuel burn-up.}
\label{fig:power_evo}
\end{center}
\end{figure}

For reference we give in Figure~\ref{fig:power_evo} 
the time evolution of the \nuebar
interaction rate expected in the \textsc{Palo Verde} detector, 
calculated as described above from the 
plant data.    This time evolution is typical of 
a plant with more than one reactor.
Refueling outages (about 1 month long each) 
give different rate excursions due to
the different distances between the reactors and the detector.    
Short accidental 
reactor trips are also visible along with the steady rate decline through the 
reactor cycle due to fuel burn-up.

\begin{figure}[h]
\begin{center}
\epsfxsize=2.7in \epsfbox{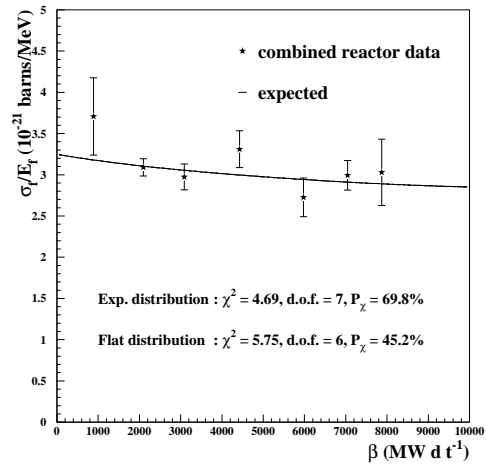}
\caption{The measured \nuebar event rate at \textsc{Chooz}, expressed as
the cross section
per fission divided by the energy per fission (see text for
explanation),
as a function of the reactor burn-up (in units of MW$\times$days per
ton), i.e., the accumulated reactor power per unit mass of fuel.
The fits to the expected rate, which
includes the reactor burn-up, and to the flat rate, where the burn-up
is neglected, are also shown. Adapted from  Nicol\`o (1999).}
\label{fig:chooz_bur}
\end{center}
\end{figure}

In Fig.~\ref{fig:chooz_bur} the measured effect of the
changing reactor fuel composition is shown. At the same time,
the figure demonstrates how small that effect really is.
We plot in  Fig.~\ref{fig:chooz_bur}
$\sigma_f/E_f$ since the number of events $n_\nu$ at a given time
and fuel composition is
\begin{equation}
n_\nu = \frac{1}{4 \pi R^2} \frac{ W_{th}}{\langle E_f \rangle} N_p
\varepsilon
\sigma_f ~,
\end{equation}
where $R$ is the distance, $N_p$ the number of protons, 
$\varepsilon$ is efficiency for the event
detection, $\sigma_f$ is the effective
cross section per fission and $\langle E_f \rangle$ 
is the average energy per fission which are both sensitive to the burn-up.

In conclusion, the \nuebar spectra, and its absolute normalization,
are known to about 2\% accuracy. Obviously, the reactors
as \nuebar sources are perfectly isotropic.
The differences between various
reactors, and the time changes due to the fuel burn-up are small,
and well understood. They do not cause additional uncertainty.
Thus, for the reactor neutrino oscillation searches at a few
percent accuracy, no
short distance `monitor' detectors are needed. One can simply
compare the measured positron spectra at the distance $L$ to
the expectation at the source ($L = 0$).

\subsection{Other reactor neutrino experiments}

We shall only briefly mention here other experiments
with reactor \nuebar that are of importance as
a potential source of information about
neutrino intrinsic properties. Besides protons, 
two targets have been extensively studied: deuterons and electrons.

Reactor \nuebar can cause deuteron disintegration by two
reaction channels; the charged current (CC) 
\begin{equation}
\bar{\nu}_e + d \rightarrow e^+ + n + n ~,
\end{equation}
with the threshold of 4.03 MeV ($M_n + m_e - M_p + B$), 
where $B$ is the deuteron binding energy,
and the neutral current (NC)  
\begin{equation}
\bar{\nu}_e + d \rightarrow \bar{\nu}_e + p + n ~,
\end{equation}
with lower threshold of $B$ = 2.23 MeV.
The cross sections of these reactions, $\sim 1.1 \times 10^{-44}$
cm$^2$/fission for the CC, and $\sim 3.1 \times 10^{-44}$
cm$^2$/fission for the NC (see e.g. Davis {\it et al.} (1979)
and Vogel  {\it et al.} (1981)) are more than
an order of magnitude smaller than for the reaction on the proton
target. (The cross section is expressed `per fission' because the
fission rate is the quantity more directly related to the reactor
power than the \nuebar flux.)

The study of the \nuebar + $d$ reactions was pioneered by
Reines and collaborators (Pasierb {\it et al.} 1979, 
Reines {\it et al.} 1980), who observed
the corresponding  two or one neutron captures. From the point of view
of neutrino oscillations, the ratio of the CC and NC rates
is potentially useful, since the CC is flavor sensitive
and NC is not (the same idea is being pursued in the case
of solar neutrinos by the SNO collaboration). More recent
experiments (Riley et al. 1999, Kozlov et al. 2000)
 show, as expected given the short
distance from the reactor, no indication of oscillations. 

The other reaction observed with the reactor \nuebar is the
scattering on electrons
\begin{equation}
\bar{\nu}_e + e \rightarrow \bar{\nu}_e + e ~,
\label{eq:nue_scatt}
\end{equation}
where the spectrum of the recoil electrons (originally assumed
at rest in the laboratory since the electron momentum
associated with the atomic binding is usually negligible) is observed. Obviously,
the reaction signature, just the recoiling electron, is
quite difficult to distinguish from background caused by 
radioactivity, making the observation of the
\nuebar$ -~e$ scattering very challenging.

The cross section for \nuebar$-~ e$ scattering consists of the well
understood weak interaction part, and a so far unobserved incoherent
electromagnetic part:
\begin{eqnarray}
\frac{{\rm d}\sigma}{{\rm d}T}  & =  & \frac{G_F^2 m_e}{2\pi}
\left[ (f + g)^2  +  (f - g)^2 \times  
 \left( 1 - \frac{T}{E_{\nu}} \right)^2 \right. \nonumber  \\
& + & \left. (g^2 - f^2) \frac{m_e T}{E_{\nu}^2} \right]   
 +   \frac{\pi \alpha^2 \mu_{\nu}^2}{m_e^2} \frac{1 - T/E_{\nu}}{T} ~,
\label{eq:esc}
\end{eqnarray}
where $T$ is the kinetic energy of the recoiling electron.
The first part, weak scattering,  represents the sum of 
coherent (interfering) contributions from charged
and neutral currents, while the second part,
proportional to $\alpha^2$, can be described 
as due to a finite neutrino magnetic moment $\mu_{\nu}$.
Only massive neutrinos can have magnetic moments, and hence
the study of this reaction, and the possible determination of
$ \mu_{\nu}^2$, is of great interest.

Again, pioneering results were obtained by Reines and collaborators
(Reines, Gurr and Sobel 1976). More recently, limits on $ \mu_{\nu}^2$
with reactor neutrinos, albeit with a rather poor signal to noise
ratio, were obtained by Derbin {\it et al.} (1993).  
At present a slightly more stringent 
direct limit of  $ \mu_{\nu} \le 1.6\times10^{-10} \mu_{\rm B}$ 
comes not from the reactor neutrinos, but from the analysis of the
shape of the Super-Kamiokande (Fukuda {\it et al.}  1999)
solar neutrino data (Beacom and Vogel 1999). A new effort to improve
the sensitivity to  $ \mu_{\nu}^2$ is currently
underway (Amsler {\it et al.} 1997) at the Bugey reactor (MUNU experiment).

\section{Experiments motivated by the atmospheric neutrino anomaly}

Two experiments have been built with the specific purpose 
of testing the hypothesis
that neutrino oscillations occur with the parameters found 
by the atmospheric neutrino
measurements. Both experiment are now completed. 
All \textsc{Chooz}  data are 
published in Apollonio {\it et al.} 1998, 1999, and Nicol\`o 1999. Data-taking 
at \textsc{Palo Verde} finished in the summer
2000 and the results 
are published in Boehm {\it et al.} 2000a, 2000b, 2001, and Miller 2000.

As it can be seen from Figure~\ref{fig:rate_mass}, in order to access 
$\Delta m^2 \approx 10^{-3}$~eV$^2$ with reactor neutrinos, 
a baseline of order 1~km
and a mass in excess of a few tons are needed.  
Indeed, the backgrounds from cosmic
radiation and natural radioactivity are 
a major consideration in the design of such 
large, low-energy detectors, and different backgrounds situations 
led the two groups to rather different designs.

The \textsc{Chooz} detector was built in a pre-existing 
underground cavity under a $\simeq 100$~m rock
overburden ($\simeq 300$~m.w.e).    
This substantial cosmic radiation shielding allowed the 
use of a homogeneous detector where 
inverse-$\beta$ events were tagged as a double (delayed) 
coincidence between the $e^+$ (prompt) 
signal and the $n$ (delayed) one.  Such simple 
event signature can be identified 
with large efficiency so that a 5 ton active mass was 
sufficient for the experiment.   
The \textsc{Palo Verde} detector, on the other hand, was 
located in an underground bunker excavated for the purpose.   
Economic considerations 
limited the overburden to 12~m ($\simeq 32$~m.w.e.) 
sufficient only to eliminate the 
hadronic component of the cosmic radiation 
and reduce the muon flux by a factor of five.   
The rather large remaining muon flux 
produced a substantial quantity of secondary neutrons 
so that a segmented detector was needed
to take full advantage of the triple coincidence 
given by the $e^+$ ionization and 
subsequent $\gamma$'s from annihilation.   
This more elaborate topological signature reduced
the detector efficiency and pushed the fiducial mass to 12 ton.
Both detectors were built with materials selected for low radioactivity and 
included a passive $\gamma$ and neutron shield 
and an active cosmic-ray veto counter.

While the 3-reactor plant of Palo Verde produces 
a larger power (11.6~GW$_{th}$) than
the 2-reactor one at \textsc{Chooz} (8.5~GW$_{th}$), 
more important is the fact that, unlike in
the case of \textsc{Palo Verde} which was a plant already running for
a long time, 
the \textsc{Chooz} reactors were commissioned
after the start of data taking of the experiment.   
This endowed the collaboration with the 
rather unique opportunity of observing 
the backgrounds at reactor-off for a substantial
period and the slow ramp-up of power.       
On the other hand the need to cope with a much
more stable operation, with the periodic 
$\sim 2/3$ refueling power excursions shown in 
Figure~\ref{fig:power_evo}, motivated the \textsc{Palo Verde} 
group to develop new methods for background
subtraction that will be important 
for future experiments likely to run in
similar steady-state situations.

\begin{figure}[h]
\begin{center}
\epsfysize=3.3in \epsfbox{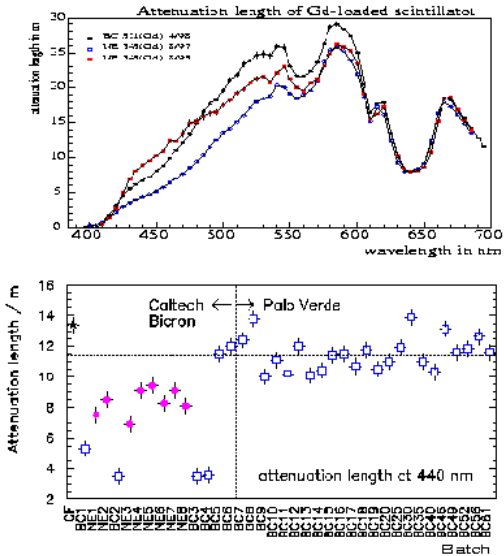}
\caption{ Initial light attenuation
length in \textsc{Palo Verde} scintillator.
Top: attenuation length as function of wavelength
in three pre-production batches.
Bottom: attenuation length at $\lambda = 440$~nm
for the initial batches of scintillator.
Production batches are to the right of the vertical dashed line,
while on the left we
show various test batches not used in the detector.
The first point (``CF'') refers to a
standard (non Gd loaded) scintillator fluid.
Typical production scintillator had an
attenuation length better than 11~m.}
\label{fig:scint_al}
\end{center}
\end{figure}

Both detectors used liquid scintillator loaded 
with 0.1\% natural gadolinium which has a high 
thermal neutron capture cross-section 
and releases a large amount of energy in the capture.
In this way the neutron capture time is reduced 
to $\sim$27~$\mu$s from $\sim$170~$\mu$s for 
the unloaded scintillator, 
proportionally reducing the uncorrelated background. 
Furthermore, Gd de-excitation after the capture 
releases a 8~MeV~$\gamma$ cascade, 
whose summed energy gives a robust event tag well above natural radioactivity. 
In contrast, neutron capture on protons 
results only in a single 2.2~MeV~$\gamma$.
While the Gd loading offers obvious advantages 
in suppressing backgrounds, it is not easy to
achieve in the stable and sufficiently transparent 
form needed for a large detector.
Both groups invested substantial resources in scintillator development.   
In Figure~\ref{fig:scint_al} we show initial attenuation length 
data for the \textsc{Palo Verde}
scintillator that was a cocktail of 60\% mineral oil, 
36\% pseudocumene (1,2,4 trimethylbenzene) 
and 4\% alcohol (used to keep the Gd compound in solution), 
with PPO as primary fluor.
This scintillator had a H:C ratio of $\simeq 2$ and 
a light yield of 56\% of anthracene, with
typical light attenuation length greater 
than 11~m at $\lambda = 440$~nm (Piepke {\it et al.} 1999).

\begin{figure}[h]
\begin{center}
\epsfysize=3.0in \epsfbox{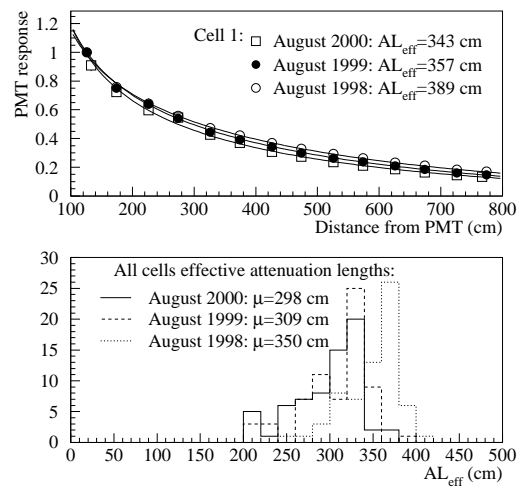}
\caption{ Evolution of scintillation-light
attenuation-length in \textsc{Palo Verde}
scintillator during the two years of life of the detector.
Top: the attenuation curve for
scintillation events at different locations along
one of the 9~m-long cells. Bottom: {\it effective}
attenuation values for all detector cells.
Note that the shorter value of the effective
attenuation length reflects the non-trivial optics of the cells.}
\label{fig:evol_al}
\end{center}
\end{figure}

The time stability of the same scintillator 
is shown in Figure~\ref{fig:evol_al} where
the light attenuation curve for one cell 
and the {\it effective} attenuation length
for all cells are presented as measured three times, 
at one year intervals. The 12\% loss
in the first year decreases to 3\% in the second, 
possibly indicating a gradual stabilization.

\begin{figure}[h]
  \begin{center}
  \epsfysize=2.3in \epsfbox{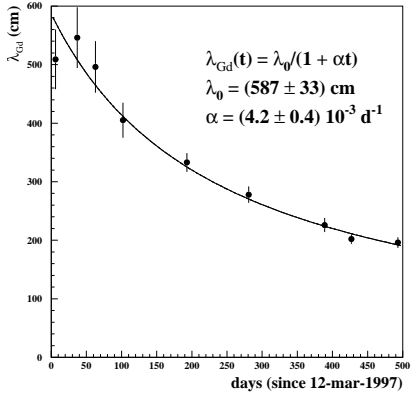}
    \caption{\small Light attenuation 
length $\lambda_{Gd}$ of the  \textsc{Chooz} Gd-loaded
scintillator  versus time with best-fit function
             shown.}
    \label{fig:tlam}
  \end{center}
\end{figure}

The time variation of the attenuation length $\lambda_{Gd}$
of the Gd-loaded scintillator  at \textsc{Chooz}, 
regularly measured in the detector, 
is well fitted by the empirical function
\begin{equation}
  \lambda_{Gd}(t) = \frac{\lambda_0}{1+\alpha t}
  \label{tlam}
\end{equation}
which accounts for the observed exponential decrease 
of the signal with time. 
Results of the fit are shown in Figure~\ref{fig:tlam}.

The properties of the Gd-loaded and unloaded 
veto liquid scintillators 
developed for \textsc{Chooz} are presented in Table~\ref{tab:scint}.

\begin{table}[htb]
  \caption{\small Main properties of the liquid scintillators
      used in the \textsc{Chooz} experiment.}
  \label{tab:scint}
  \begin{center}
    \begin{tabular}{|l|c|c|}
      & Gd-loaded & unloaded \\
      \hline
      Chem. cont.: & & \\
       ~~basic & Norpar-15\footnote{Norpar-15 is the trademark 
of Exxon Mobil Corporation.} & Mineral oil \\
      & (50\% vol.) & (92.8\% vol.)\\
       ~~aromatics,  & IPB+hexanol & IPB \\
      ~~alcohols & (50\% vol.) & (7.2\% vol.) \\
       ~~wavelength  & p-PTP+bis-MSB & PPO + DPA \\
      ~~shifters & (1 g/l) & (1.5 g/l) \\
      \hline
      Atomic mass & & \\
      ~~composition: & & \\
      \hspace{0.5cm} H & $12.2\%$ & $13.3\%$ \\
      \hspace{0.5cm} C & $84.4\%$ & $85.5\%$ \\
      \hspace{0.5cm} Gd & $0.1\%$ & \\
      \hspace{0.5cm} others & $3.3\%$ & $1.2\%$ \\
      \hline
      compatibility & \multicolumn{2}{c}{acrylic, teflon} \vline \\
      \hline
      density (20$^{\circ}$) & 0.846 g/l & 0.854 g/l \\
      \hline
      Flash point & 69$^{\circ}$ & 110$^{\circ}$ \\
      \hline
      Scint. yield &\multicolumn{2}{c}{5300 photons/MeV
        ($\simeq 35\%$ of anthracene)} \vline \\ 
      \hline
      Optical  & 4 m & 10 m \\
      ~~atten. length  & & \\
      \hline
      Refr. index  & 1.472 & 1.476 \\  
      \hline
      Neutron  & 30.5\usec & 180\usec \\
      ~~capture  time  & - & \\ 
      \hline  
      Thermal neutron & $\sim$ 6 cm & $\sim$ 40 cm \\
      ~~capture path  & & \\
      \hline
      Capture  & $84.1\%$ & \\
      ~~fraction on Gd  & & \\
      \end{tabular}
  \end{center}
\end{table}

\subsection{\textsc{Chooz}}

\begin{figure*}[ptb]
\begin{center}
\epsfysize=3.8in \epsfbox{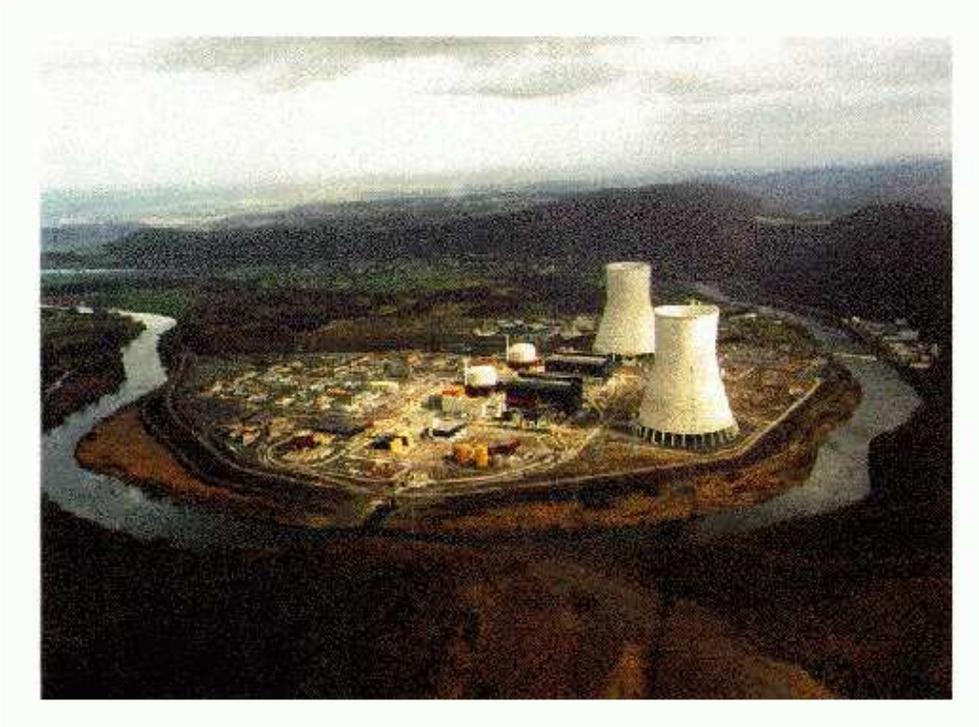}
\vspace{0.5cm}
\caption{ Aerial view of the \textsc{Chooz} power plant.
The detector is located in a tunnel
under the hills on the bottom right of the photograph.}
\label{fig:chooz_view}
\end{center}
\end{figure*}

The \textsc{Chooz} detector was built at the distances of 
1115 m and 998 m from the two reactors of the new \textsc{Chooz}
power plant of {\it \'Electricit\'e de France} in 
the Ardenne region of France.   The plant,
shown in Figure~\ref{fig:chooz_view}, 
has a total thermal power of 8.5~GW$_{th}$ and the two
reactors reached full power in, respectively, May and August 1997.   
The experiment took
data from April 1997 until July 1998, 
in the conditions specified in Table~\ref{tab:chooz_data}.

\begin{figure}[h]
\begin{center}
\epsfysize=3.4in \epsfbox{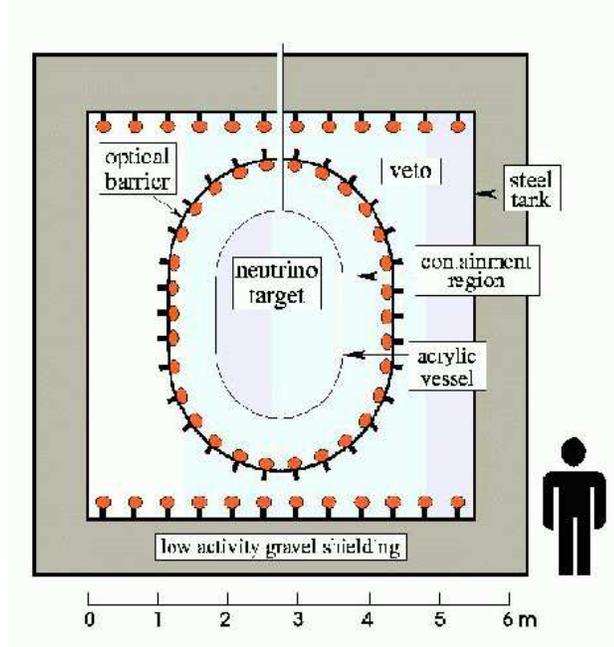}
\caption{ Schematic drawing of the \textsc{Chooz} detector.}
\label{fig:choozdet}
\end{center}
\end{figure}

The apparatus, schematically shown in Figure~\ref{fig:choozdet}, 
consisted of a central volume
of scintillator with a mass of 5 tons, 
where {\nuebar} were detected.
This scintillator was contained in an acrylic vessel 
(region 1) that separated it from a 70~cm thick
shielding layer of mineral oil (region 2).   
192 eight-inch photomultipliers (PMT's) were mounted onto
a steel vessel that, in-turn, isolated, 
mechanically 
and optically, the central detector
from the outer veto counter.    The central detector 
had a photocathode coverage of 15\%
and a light yield of $\sim 130$~photoelectrons/MeV 
(p.e./MeV) (Baldini {\it et al.} 1996).
The 90~ton veto scintillator was at least 
80~cm thick and was readout with two rings of
24 eight-inch PMT's, the outer containment 
tank being painted with white reflective paint.
An outer layer (75~cm thick) of low-activity 
sand provided primary shielding from the
rock.  

 Laser flashers were installed to 
monitor the detector performance and radioactive
sources could be inserted 
into the central region 
of the detector through special pipes.
The detector energy response was calibrated 
daily with $^{60}$Co, $^{252}$Cf and AmBe
$\gamma$- and n-sources in order to accurately 
track the aging of the scintillator,
the detector efficiency and the energy calibration.

\begin{figure}[h]
\begin{center}
\epsfysize=3.3in \epsfbox{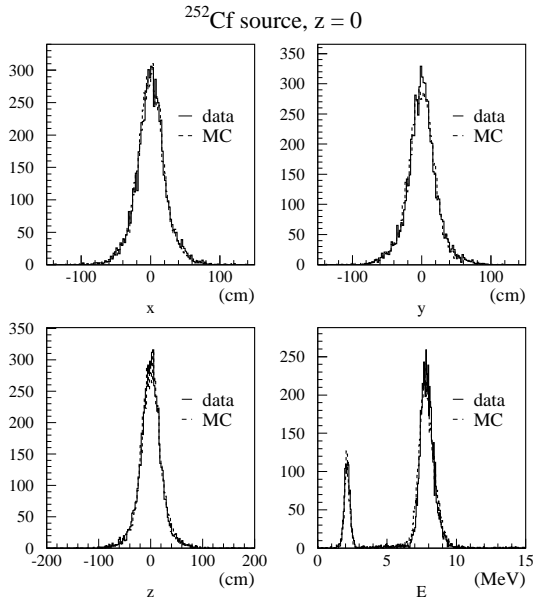}
\caption{ Visible energy and position reconstructed for a calibration
$^{252}$Cf source placed inside the \textsc{Chooz} detector.}
\label{fig:chooz_Cf}
\end{center}
\end{figure}

As an example we show in Figure~\ref{fig:chooz_Cf} 
the results of a Cf calibration run
with the source placed in the middle of the detector.    
Data is compared with a Monte
Carlo simulation for the reconstruction of 
the $x$, $y$ and $z$ positions and total
energy in the detector.  Both the peaks 
for n-captures on $p$ (2.2~MeV) and Gd (8~MeV)
are clearly visible.   The very good energy 
resolution ($\sigma(E)/E = 5.6~\%$
at 8~MeV) allows one to verify that 
the 8~MeV peak is in fact the superposition of a
7.77~MeV line with 77\% weighting 
from capture on $^{157}$Gd (energy shifted from 

\begin{table}[hb]
\caption{Summary of the \textsc{Chooz} data-taking conditions.}
\begin{center}
\begin{tabular}{|lcc|}
                       & Time (h) & $\int{W_{th} dt}$~(GWh$_{th}$) \\
\hline
Total run              &  8761.7  &                                \\
Live                   &  8209.3  &                                \\
Dead                   &   552.4  &                                \\
Reactor 1 ON only      &  2058.0  &       8295                     \\
Reactor 2 ON only      &  1187.8  &       4136                     \\
Both reactors ON       &  1543.1  &       8841                     \\
Both reactors OFF      &  3420.4  &          0                     \\
\end{tabular} 
\label{tab:chooz_data}
\end{center}
\end{table}

\noindent
7.94~MeV because of scintillator saturation effects) 
and a 8.31~MeV line with
23\% weighting from capture on $^{155}$Gd 
 (energy shifted from 8.54~MeV).   
The fit to these two Gaussians gives
$\chi^2/{\rm d.o.f.} = 67.6/55$ while the fit to 
a single Gaussian is very poor with
$\chi^2/{\rm d.o.f.} = 875/58$.  
The position resolution was found to be 
$\sigma_x = \sigma_y = \sigma_z = 17.5$~cm.
\indent

\subsection{\textsc{Palo Verde}}

\begin{figure}[htb]
\begin{center}
\epsfysize=3.8in \epsfbox{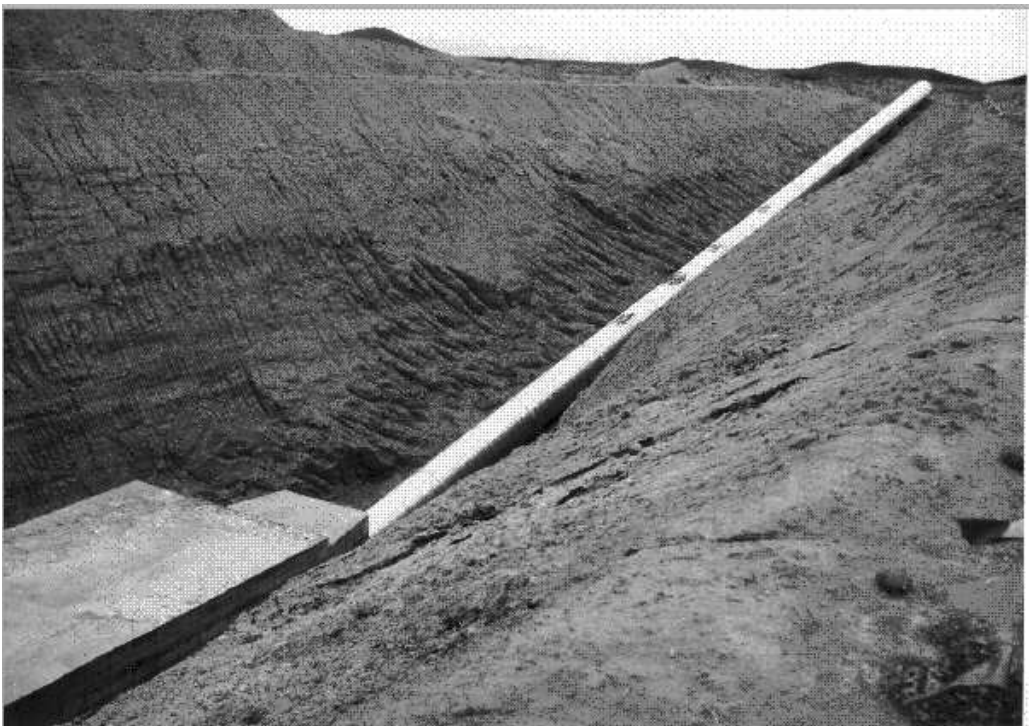}
\caption{\footnotesize The \textsc{Palo Verde} 
underground laboratory at the time of
construction (Fall 1996).}
\label{fig:PV_bunker}
\end{center}
\end{figure}

The \textsc{Palo Verde} 
experiment was built at the Palo Verde Nuclear Generating 
Station, the largest nuclear plant in the Americas, 
$\sim 80$~km west of Phoenix, 
in the Arizona desert. The total thermal power 
from three identical pressurized 
water reactors is 11.6~GW$_{th}$.      
Two of the reactors were located 890~m from the 
detector, while the third was at 750~m. The shallow underground bunker housing 
the detector is shown in Figure~\ref{fig:PV_bunker} 
at the time of construction.
In total 350.5~days of \nuebar data were collected at 
\textsc{Palo Verde} in the period
between October 1998 and July 2000, covering 
four scheduled refueling outages as indicated
in Figure~\ref{fig:power_evo}. Of these,  
242.2~days were at full power, while 21.8~days
(86.5~days) had the reactor at 750~m (890~m) off.
Such data was complemented by frequent calibration runs.

\begin{figure*}[htb]
\begin{center}
\epsfysize=3.5in \epsfbox{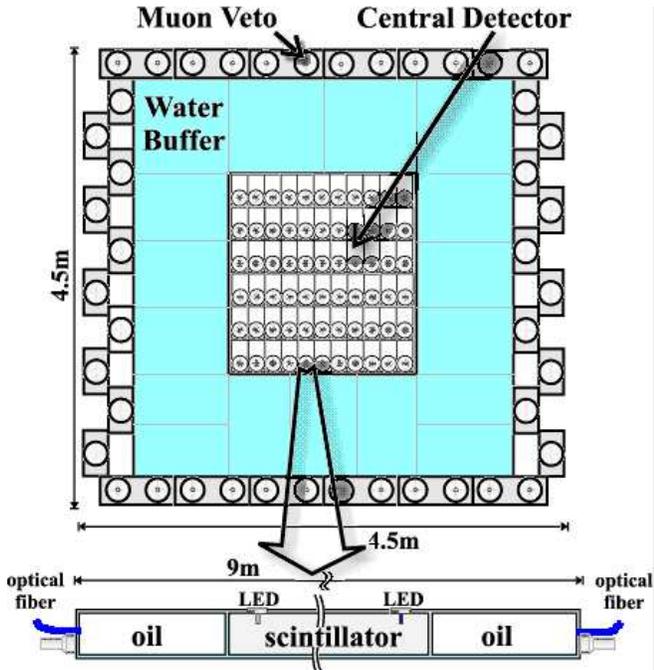}
\caption{ Schematic view of the \textsc{Palo Verde} neutrino detector.}
\label{fig:detector}
\end{center}
\end{figure*}

The fiducial mass, segmented for active background rejection, consisted of 
66 acrylic tanks filled with 0.1\% Gd-loaded scintillator and arranged 
as shown in Figure~\ref{fig:detector}.  Each cell was 9~m long, with a 
$12.7\times 25.4$~cm$^2$ cross section, and it was viewed by two 5-inch 
photomultiplier tubes, one at each end.
A $\bar{\nu}_e$ is identified by space- and time-correlated 
$e^+$ and $n$ signals.
Positrons deposit their energy in a scintillator cell and annihilate,
yielding two 511~keV $\gamma$'s that, in general, will be detected in
different cells, giving a triple coincidence. Neutrons thermalize and  
are captured in Gd, giving a $\gamma$-ray shower of 8~MeV total energy also
detected in more than one cell. 
The central detector was surrounded by a 1~m-thick water shield to moderate 
background neutrons produced by muons outside the detector 
and to absorb $\gamma$'s from the laboratory walls.
Outside of the water tanks were 32 large liquid scintillator counters
and two end-caps to veto cosmic muons. The  rate of cosmic muons was 
approximately 2~kHz. The pattern of muons traveling through 
veto counters and their timing relative to the central detector 
hits were recorded for subsequent off-line analysis.
The central detector was equipped with a system of 
tubes that allowed the insertion
of calibration sources in the small spaces between cells.   
In addition, a set of blue LEDs and optical fibers 
was used to produce flashes of light 
inside each of the cells.
In order to reduce natural radioactivity, all building materials for the 
detector were carefully selected, including the aggregate (marble) used in the
concrete of the underground  laboratory.

Both the positron and the neutron 
were triggered by a triple coincidence requiring 
at least one cell above a ``high'' threshold set at $\sim$600~keV (positron 
ionization or neutron capture cascade), 
and two cells above a ``low'' threshold set 
at $\sim$40~keV (Compton scattering from 
annihilation photons or neutron capture cascade 
tails). The triple coincidences were 
required to be within $3\times 5$ matrices of cells 
anywhere in the detector as recognized 
by a custom-made trigger processor (Gratta {\it et al.} 1997).

The efficiency calibration was based upon 
a primary measurement performed a few 
times per year with a calibrated $^{22}$Na 
$e^+$ source and an Am-Be neutron 
source.  The $^{22}$Na source  mimicked 
the effects of the positron from the $\rm \bar\nu_e$ interaction by providing 
annihilation radiation and a 1.275~MeV photon which simulated the $e^+$ 
ionization in the scintillator.   The source was placed at 18 positions in the 
detector deemed to be representative of different conditions. 
The results of this 
procedure were then rescaled to the $e^+$ case using a Monte Carlo simulation.
The neutron detection efficiency was measured by scanning the detector with the
Am-Be source where the 4.4~MeV $\gamma$ associated 
with the neutron emission was 
tagged with a miniaturized NaI(Tl) counter.  
Other calibrations, used to measure the detector energy response,  were 
performed using the Compton edges from $^{137}$Cs, $^{65}$Zn and $^{228}$Th
sources.       The same Th source was also used more frequently to track the
scintillator transparency, as already shown in Figure~\ref{fig:evol_al}.
Weekly runs of the fiber-optic and LED flasher systems were used, respectively,
to monitor the gain and linearity of photomultipliers and the timing/position 
relationship along the cells.    

\begin{figure}[h]
\begin{center}
\epsfysize=3.5in \epsfbox{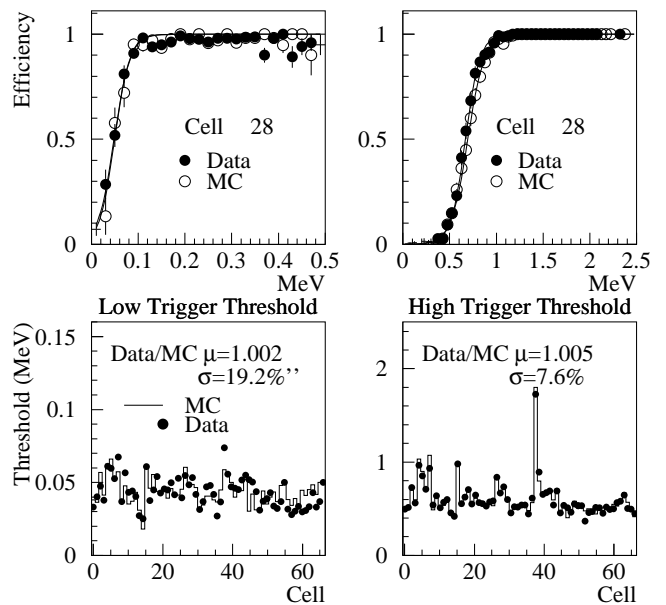}
\caption{ A comparison of the trigger thresholds 
at \textsc{Palo Verde} from data and Monte Carlo.
The data were taken with a $^{22}$Na source at the center of each cell.
The top portion shows the efficiency
of the trigger thresholds (low and high) for a typical
cell as a function of energy deposited;
the bottom shows the energy at 50\% efficiency
for low and high thresholds in all 66 cells.}
\label{fig:trig_thr}
\end{center}
\end{figure}

Since the  energy deposition of the 511~keV $\gamma$'s in one cell 
has a sharp falling spectrum (Compton scattering) it was vital to have 
the lowest possible ``low'' thresholds in the trigger and to understand 
the behavior of  such thresholds with great accuracy.
This second task was complicated by the fact that the trigger used 
voltage amplitudes, while only charge from 
integrating ADCs was available off-line.
For this reason the detector simulation included a  detailed  
description of the signal development in time.
This code correctly described the shape of pulses taking into 
account scintillator light yield, attenuation length and 
de-excitation time; photomultiplier rise- and fall-time and gain; and event
position along a cell.
The simulation of the detector response to the $^{22}$Na source 
is shown in Figure~\ref{fig:trig_thr} and correctly
describes the 40~keV (600~keV) threshold position to within 1.4~keV (2.6~keV),
resulting in  uncertainties on the positron and neutron efficiencies 
of 4\% and, respectively, 3\%. 

\subsection{Backgrounds}

\begin{figure*}[h]
\begin{center}
\epsfysize=2.1in \epsfbox{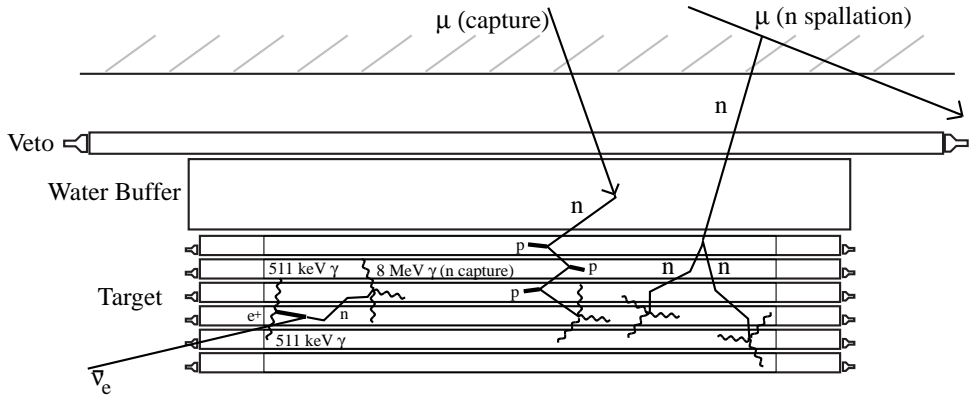}
\caption{ Schematic view of two types of cosmic-$\mu$-induced backgrounds
and a signal event (far left) in the \textsc{Palo Verde} detector.
Neutrons are produced by cosmic-$\mu$
spallation (right) and capture (left) on the
materials outside the veto counter.
Both can result in either
neutron thermalization and capture (left),
where the thermalization process fakes
the prompt triple coincidence, or secondary neutrons
production (right), where one of the
captures fakes the prompt triple coincidence.}
\label{fig:n_backgnd}
\end{center}
\end{figure*}

There are generally two types of background 
affecting long baseline reactor experiments
where the signal is based on the correlated
$e^+, n$ signature:
uncorrelated hits from cosmic-rays and 
natural radioactivity and correlated ones 
from cosmic-$\mu$-induced neutrons.   
The first type can be measured by studying the 
time difference between positron-like and neutron-like parts of an event. 
More insidious are  cosmic-$\mu$-induced 
neutrons that present the same time and space correlation
between prompt and delayed parts of the event as in \nuebar.  Such events are 
schematically shown in the 
\textsc{Palo Verde} detector, in Figure~\ref{fig:n_backgnd}.
Neutrons are produced by cosmic-$\mu$ spallation and capture on the 
materials outside the veto counter.   
Both production mechanisms can result in either
neutron thermalization and capture, where the thermalization process fakes
the prompt triple coincidence, 
or secondary neutrons production, where one of the 
captures fakes the prompt triple coincidence.   
Conceptually the same situation holds
for \textsc{Chooz} (and \textsc{Kamland}, as it will be discussed later), 
although differences in overburden
and the simpler scheme of coincidence, 
numerically change the relative importance of
different backgrounds.   
It is useful to point out that direct neutrons from the reactors have a totally
negligible effect at the distances discussed here.

\begin{figure*}[h]
\begin{center}
\epsfysize=2.4in \epsfbox{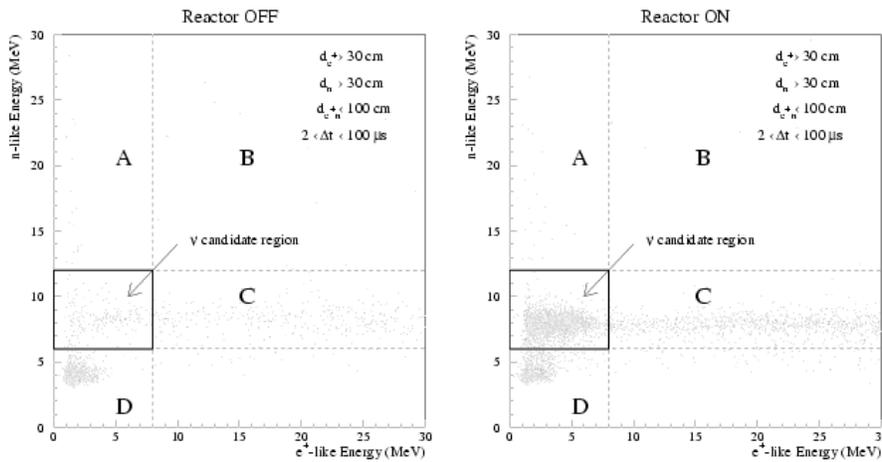}
\caption{ Delayed energy vs. prompt energy in pre-selected \textsc{Chooz} 
events.
The selection cuts are listed in the figure.
On the left it is shown the case of reactors
OFF, while on the right is the case of reactors ON.
A description of the event-types
in the different regions of the plot is given in the text.}
\label{fig:Chooz_scatter}
\end{center}
\end{figure*}

Both experiments pre-select \nuebar candidates 
by requiring an appropriate topology 
(in space and time) for the prompt and delayed parts of each event and their 
relative position.   Such cuts insure that 
the spatial and temporal extents of the 
events are compatible with the \nuebar 
hypothesis and that events are well contained and 
measured in the detector.   
A general classification in terms of signal and 
different backgrounds can be conveniently
done by studying the correlation between prompt 
and delayed energy in \textsc{Chooz} for such 
pre-selected sample, as shown in Figure~\ref{fig:Chooz_scatter}.   
The region marked ``B'' in the Figure 
contains cosmic-ray muons stopping in the 
detector after entering undetected by the veto counter.  
Both prompt energy (muon 
ionization) and delayed energy (Michel electron) are large.   
Indeed events in region 
``B'' show a fast time correlation between prompt 
and delayed part, consistent with 
the muon lifetime.
Region ``C'' is populated by the muon-spallation events discussed above: large 
prompt energy deposit from proton recoils in neutron thermalization is
accompanied by a fix 8~MeV energy deposit characteristic of neutron capture.
Regions ``A'' and ``D'' are populated 
by random coincidences of natural radioactivity 
hits, sometimes including a high-energy 
proton recoil from neutron scattering in the
delayed part (region ``A'').   
Neutrino candidates populate the region framed 
by the darker line, as it can be seen
by comparison between the scatter plots with reactors ON and OFF.

\begin{figure}[h]
\begin{center}
\epsfysize=3.5in \epsfbox{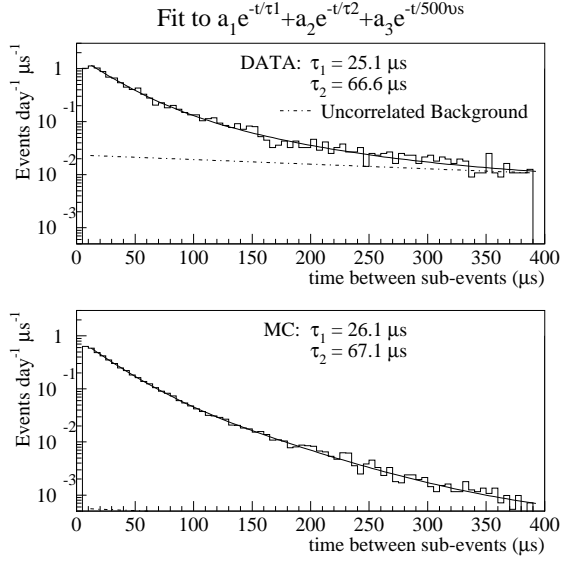}
\caption{ Time elapsed between the prompt and delayed parts of
events in \textsc{Palo Verde} data and Monte Carlo.    
The simulated data are fit to
two exponentials.  Real  data are fit to three exponentials of which the third
accounts for the random background.}
\label{fig:iet}
\end{center}
\end{figure}

The time elapsed between the prompt 
and delayed parts of the events is shown in 
Figure~\ref{fig:iet} for the \textsc{Palo Verde} data.   
We note that the process of $n$-capture
in the segmented detector requires 
the sum of two exponentials to fit properly.
This is due to the fact that a fraction 
of the neutrons stop, after thermalization,
in passive materials (mainly acrylic for \textsc{Palo Verde}) 
where there is no Gd and the 
capture is a slower process.  While the Monte Carlo gives a good fit with two 
exponentials, for data a third exponential, 
with longer time constant, is needed 
in the fit.   Such exponential accounts for 
events initiated by  uncorrelated 
background, having the delayed part triggered by cosmic rays, crossing the 
detector with a 2~kHz rate.
Timing cuts are applied by both experiments 
to insure that events are consistent
with a neutron capture.   In addition events are rejected for a period of
time following tracks detected in the veto counters.   
This last cut is particularly
important at \textsc{Palo Verde} where the cosmic-ray rate is high.

The availability for \textsc{Chooz}  of data at zero power 
and with the reactors ramping up provides
an independent way to check the magnitude of signal and background.
The fitting procedure proceeds as follows. 
For each run the predicted number of neutrino candidates 
resulting from the sum of a signal term, 
linearly depending on the reactor effective power $W^*$ , 
and the background, assumed to be constant and independent of
power \footnote{The ``effective'' power $W^*$ is a fictitious   
thermal power corresponding to both reactors
located at the reactor 1 site, and thus providing $9.55$ GW at
full operating conditions and at starting of reactor operation.}; so,
\begin{equation}
  \overline{N}_i = (B + W_i^\ast X) \Delta t_i,
  \label{nrun2}
\end{equation}
where the index $i$ labels the run number, 
$\Delta t_i$ is the corresponding live time, 
$B$ is the background rate and $X$ is the positron 
yield per unit power averaged over the two reactors.

\begin{figure}[h]
\begin{center}
\epsfysize=3.6in \epsfbox{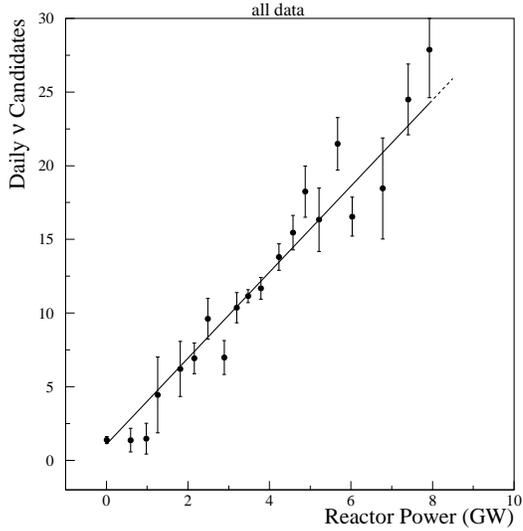}
\caption{ \textsc{Chooz} \nuebar rate during the reactors commissioning.  
The background
at reactors off is  $1.1\pm 0.25$~events/day.}
\label{fig:Chooz_turn_on}
\end{center}
\end{figure}

The results are listed in Table~\ref{tab:poilik}, 
for three data taking periods corresponding to threshold readjustments.
The data are also shown in a compact form in
Figure~\ref{fig:Chooz_turn_on}.   

\begin{figure}[h]
\begin{center}
\epsfysize=4.5in \epsfbox{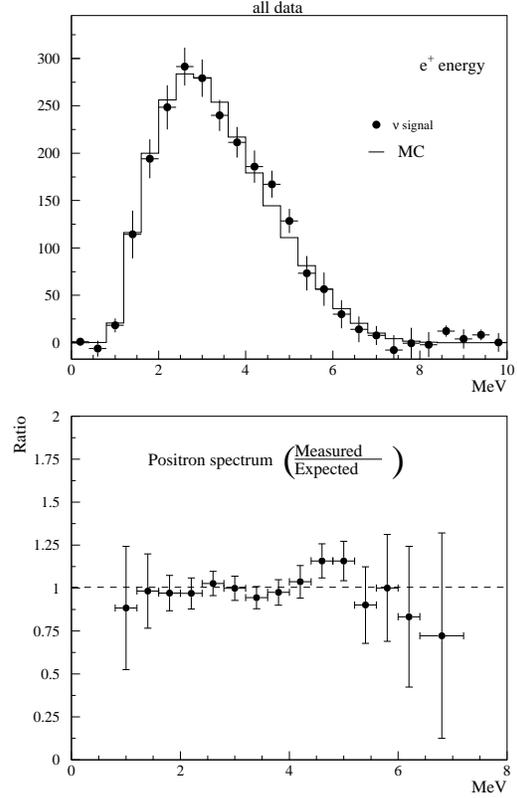}
\caption{ Background-subtracted positron energy spectrum in 
\textsc{Chooz} data.
Error bars represent statistical errors only.
The solid histogram represents the expectation
for the case of no oscillations.
The ratio between the two curves is shown on the bottom
panel.}
\label{fig:Chooz_spec}
\end{center}
\end{figure}

A simple subtraction of the $e^+$ spectra 
with reactor ON and OFF gives for \textsc{Chooz}
the spectrum shown in Figure~\ref{fig:Chooz_spec}.
The comparison of the observed distribution 
with the expected one for no-oscillations
already shows very good agreement.
\begin{table}[htb]
  \caption{ Summary of the likelihood fit 
parameters for the three data taking periods
at \textsc{Chooz} .}
  \label{tab:poilik}
  \begin{center}
    \begin{tabular}{|l|c|c|c|}
      period & 1 & 2 & 3 \\
      \hline
      \hline
      starting date   & 97/4/7 & 97/7/30 & 98/1/12 \\
      \hline
      runs   & $579 \rightarrow 1074$ & $1082 \rightarrow 1775$ &
               $1778 \rightarrow 2567$ \\
      \hline
      live time (h) & $1831.3$ & $2938.8$ & $3268.4$ \\
      \hline
      reactor-off time (h) & $38.9$ & $539.5$ & $2737.2$ \\
      \hline
      $\int W {\rm d}t$ (GWh) & $7798$ & $10636$ & $2838$\\
      \hline
      $B$ (counts/d)& $1.25\pm 0.6$ & 
$1.22\pm 0.21$ &$2.2\pm 0.14$\\
      \hline
      $X$ (counts/d GW) & $2.60 \pm 0.17$ & $2.60 \pm 0.09$ &
                                          $2.51 \pm 0.17$ \\
      \hline
      $\chi^2/dof$ & $136/117$ & $135/154$ & $168/184$ \\
      \hline
      $N_\nu$ (counts/d) & $24.8 \pm 1.6$ & $24.8 \pm 0.9$ &
                                      $24.0 \pm 1.6$ \\
      (@full power) & & &\\
    \end{tabular}
  \end{center}
\end{table}

\onecolumn

\newpage

\begin{table*}
\caption{ Summary of results from the \textsc{Palo Verde} experiment
(Boehm {\it et al.} 2001). Uncertainties are statistical only.
$N$, $N'$ and $(1 - \epsilon_1)B_{\rm{pn}}$ are measured rates,
while  Background and $R_\nu$ are efficiency corrected (assuming that the
background events are measured with the same efficiency as the signal).
$R_{\rm calc}$ is for no oscillation hypothesis.
See text for notation.}
\label{tab:results}
\begin{tabular}{lcccccccc}
Period  & \multicolumn{2}{c}{1998} & \multicolumn{2}{c}{1999-I} &
 \multicolumn{2}{c}{1999-II} & \multicolumn{2}{c}{2000} \\
Reactor & on & 890 m off & on & 750 m off & on & 890 m off & on & 890 m
off \\
\hline
Duration (d)      & 30.4 & 29.4 & 68.2 & 21.8 & 60.4 & 29.6 & 83.2 & 27.5
\\
efficiency (\%)  & 8.0  & 8.0 & 11.5 & 11.6 & 11.6 & 11.6 & 10.9 & 10.8 \\
\hline
$N$ (d$^{-1}$)  & 39.6 $\pm$ 1.1 & 34.8 $\pm$ 1.1 & 54.9 $\pm$ 0.9 &
45.1 $\pm$ 1.4
 & 54.2 $\pm$ 0.9 & 49.4 $\pm$ 1.3 & 52.9 $\pm$ 0.8 & 43.1 $\pm$ 1.3   \\
$N'$ (d$^{-1}$) & 25.1  $\pm$ 0.9 & 21.8 $\pm$ 0.9 & 33.4 $\pm$ 0.7 &
32.0 $\pm$ 1.2
 & 32.5 $\pm$ 0.7 & 32.6 $\pm$ 1.0 & 30.2 $\pm$ 0.6 & 30.4 $\pm$ 1.1  \\
$(1 - \epsilon_1)B_{\rm{pn}}$ (d$^{-1}$) & 0.88 & 0.89 & 1.11 & 1.11 &
1.11 & 1.11 & 1.07 & 1.07  \\
\hline
Background (d$^{-1}$)   & 292 $\pm$ 11 & 255 $\pm$ 10 & 265 $\pm$ 6 & 266
$\pm$ 10
 & 256 $\pm$ 6 & 265 $\pm$ 9 & 249 $\pm$ 5 & 272 $\pm$ 9 \\
$R_\nu$   (d$^{-1}$)        & 202 $\pm$ 19 & 182 $\pm$ 18 & 212 $\pm$ 10 &
124 $\pm$ 17
 & 214 $\pm$ 11 & 161  $\pm$ 15 & 237 $\pm$ 10 & 129 $\pm$ 16 \\
$R_{\rm calc}$ (d$^{-1}$)     & 216 & 154 & 218 & 129 & 220 & 155 & 218 &
154 \\
\end{tabular}
\end{table*}

\twocolumn[\hsize\textwidth\columnwidth\hsize\csname
@twocolumnfalse\endcsname]

The same procedure can be repeated for 
\textsc{Palo Verde} using the thermal power excursions
due to refueling.  However in this case this 
technique substantially magnifies the
errors since: 1) the periods of low power still 
have about 2/3 of the full flux, so
that in the subtraction most of the signal is lost, 
2) the statistical errors are
dominated by the relatively short periods of low power.   
In addition, for any experiment,
the background subtraction method will give correct
result only if special attention is paid to the
data quality, guaranteeing in particular that
the efficiencies for signal 
{\it and} background are accurately known
and remain as  constant as possible
through the experiment.

An alternative method (Wang {\it et al.} 2000) was developed 
for the \textsc{Palo Verde} analysis starting
from the evidence that, for their depth and 
detector configuration, the dominant 
correlated background has at least two neutrons, 
each triggering the detector with 
its capture.   Such intrinsic symmetry can be 
used to cancel most of the background 
directly from data and compute 
the remaining components from Monte Carlo simulations.   
This technique makes the best possible use 
of the statistical power of all data collected.
The rate of candidate events after all cuts can be written as 
$N = B_{\rm unc} + B_{\rm nn} + B_{\rm pn} + S_{\nu}$ where 
the contribution of the uncorrelated 
$B_{\rm unc}$, two-neutron $B_{\rm nn}$ and 
other correlated backgrounds $B_{\rm pn}$ are explicitly represented, along 
with the $\bar\nu_{\rm e}$ signal $S_{\nu}$.   
The dominant background $B_{\rm nn}$ (along with $B_{\rm unc}$)
is symmetric under exchange of sub-events, so that an event selection with the 
requirements for the prompt and delayed event parts swapped, 
will result in a rate 
$N^{\prime} = B_{\rm unc} + B_{\rm nn} + 
\epsilon_1 B_{\rm pn} + \epsilon_2 S_{\nu}$.
Here $\epsilon_1$ and $\epsilon_2$ account for the different efficiency for 
selecting asymmetric events after the swap.
One can then calculate $N - N^{\prime} = 
(1-\epsilon_1)B_{\rm pn} + (1-\epsilon_2)S_{\nu}$
where the efficiency correction $\epsilon_2\simeq 0.2$ can be estimated from 
the $\rm\bar\nu_e$ Monte Carlo simulation.

The \textsc{Palo Verde} group found that the processes 
of $\mu$-spallation in the laboratory 
walls and capture of the $\mu$'s that are not tagged by the veto counter, 
contribute to $(1-\epsilon_1) B_{\rm pn}$, 
while other backgrounds are negligible.
Using Monte Carlo simulation, 
they obtain $(1-\epsilon_1)B_{\rm pn}=-0.9\pm 0.5$~d$^{-1}$
($-1.3\pm 0.6$~d$^{-1}$) for $\mu$-spallation in 
the 1998 (1999) data-set; the same figures
for $\mu$-capture are $0.6\pm 0.3$~d$^{-1}$ 
($0.9\pm 0.5$~d$^{-1}$) in 1998 (1999).
This represents only a small correction 
to $N-N^{\prime}$ since the error on $B_{\rm pn}$
is reduced by the fact that
$\epsilon_1$ is close to 1.  
While the Monte Carlo model is accurate for the capture
process, in the case of spallation the broad range of spectral indexes for 
the n-recoil energy reported in literature was simulated
(Wang {\it et al.} 2000).  
The average between different predictions 
is then used for $B_{\rm pn}$ while the 
spread is used as an extra systematic error.
Since no $\bar\nu_{\rm e}$ signal is present 
above 10~MeV, the observed integrated
rate above such energy is used as a normalization of the Monte Carlo. 
The rate of neutrons produced by muon spallation has been measured
at  \textsc{Palo Verde} (Boehm {\it et al.} 2000c) and the dependence
of the neutron spallation yield on depth has been analyzed
by Wang {\it et al.} (2001).

The \textsc{Palo Verde} results obtained 
in this way are shown in Table~\ref{tab:results} for 
different running periods.     
Clearly also in this case there is good agreement 
with the no-oscillation hypothesis.

\subsection{Event reconstruction techniques}
\label{sec:reco}

The identification of the neutrino signal and
the rejection of the  background in reactor neutrino
experiments depends on the accuracy of the event energy
and position determinations and on the spatial and time
correlations of the detected positron and neutron.
Event reconstruction in a segmented detector like \textsc{Palo Verde}
is relatively simple. However, event characterization in
single vessel detectors, like \textsc{Chooz} or \textsc{Kamland},
which are viewed by photomultipliers placed at the vessel
surface, requires refined reconstruction methods
making the best use of the PMT charge and time information.
The relative importance of the time and charge information
in optimizing the detector spatial resolution
(which also affects the energy resolution) depends on several
factors: the size of the vessel, the distance between PMTs
and the fiducial target volume,
the scintillator light yield and attenuation length.
In \textsc{Chooz}, the relatively small detector volume
and the small distance between the neutrino target and the PMTs,
made the charge information the dominant one in the
precise event characterization. In the case of larger volume
detectors, filled with non Gd loaded scintillator,
like \textsc{Kamland} (Alivisatos {\it et al.} 1998), 
or also,
like the solar neutrino experiment \textsc{Borexino}
(Alimonti {\it et al.} 1998, 2000),
the time information gains importance.

As an example of a minimization algorithm for event reconstruction,
based on the PMT measured charge, we describe the one used for
\textsc{Chooz}.
The standard algorithm uses a maximum likelihood method
to reconstruct the energy $E$ and the vertex $\vec{x}$ of an event.
The likelihood is defined as the joint Poissonian
probability of observing a measured distribution of photoelectrons
over 24 ``patches'', each grouping 8 adjacent PMTs,
for given $(E,\vec{x})$ coordinates in the detector.
So, for an event occurring at time $t$ after
the start of data taking, one can build a likelihood function as
follows:
\begin{equation}
  {\cal L}(N;\overline{N}) =
             \prod_{j=1}^{24} P(N_j;\overline{N}_j(E,\vec{x},t)) =
             \prod_{j=1}^{24} \frac{\overline{N}_j^{N_j}}{N_j!}
             e^{-\overline{N}_j} ~,
  \label{nphelike}
\end{equation}
where $N_j$ is the observed number of photoelectrons and
$\overline{N}_j$ the
expected one for the j-th patch given an event $(E,\vec{x},t)$. The
reason for
using a Poissonian instead of Gaussian statistics is due to the frequent
occurrence of low energy events with low number of photoelectrons
detected by
some PMT patches.
The predicted number of photoelectrons for each patch is computed by
considering
a local deposit of energy, resulting in a number of visible photons
which are
tracked to each PMT through the different attenuating Region 1
(Gd-doped) and
Region 2 scintillators. Therefore
\begin{equation}
  \overline{N}_j = \alpha E \eta \sum_{k=1}^8
\frac{\Omega_{jk}(\vec{x})}{4\pi}
    \exp\left(-\frac{d_1^{jk}(\vec{x})}{\lambda_{Gd}(t)}
         -\frac{d_2^{jk}(\vec{x})}{\lambda_{Hi}}\right)
  \label{nthe}
\end{equation}
where
$E$  is the ionization energy deposited in the scintillators,
$\alpha$  is the light yield of the scintillator,
$\eta$  is the average PMT quantum efficiency,
$\Omega_{jk}$  is the solid angle subtended by the k-th PMT from the
event position,
$d_1^{jk}$  is the light path length in region 1,
$d_2^{jk}$  is the light path length in region 2,
$\lambda_{Gd}$  is the attenuation length in region 1 scintillator,
and $\lambda_{Hi}$  is the attenuation length in region 2  scintillator.

The solid angle is approximated by the  expression
\begin{equation}
  \Omega_{jk} = 2\pi \left(1 - \frac{d_{jk}}{\sqrt{d_{jk}^2+r_{PMT}^2
\cos\theta}}\right)\,
  \label{solid}
\end{equation}
$r_{PMT}$ being the PMT photocathode radius, $d_{jk}=d_1^{jk}+
d_2^{jk}$, and $\theta$ being the angle between the
event-PMT direction and the inward unit vector normal to the PMT
surface.

Instead of (\ref{nphelike}), as is usually the case for problems
involving the
maximum likelihood method, it is more convenient to use the theorem on
the
``likelihood ratio test'' for goodness-of-fit to convert the likelihood
function into the form of a general $\chi^2$ statistics (Eadie {\it et al.} 1971).
If one assumes ${N_j}$ to be the best estimate of the true (unknown)
photoelectron distribution
and can form the likelihood ratio $\lambda$ defined by
\begin{equation}
  \lambda = \frac{{\cal L}(N;\overline{N})}{{\cal L}(N;N)}
  \label{likerat}
\end{equation}
The ``likelihood ratio test'' theorem states that the ``Poissonian''
$\chi^2$,
defined by
\begin{equation}
  \chi^2 = -2 \log \lambda = 2 \sum_{j=1}^{24} [ \overline{N}_j - N_j +
N_j
           \log(\frac{N_j}{\overline{N}_j})],
  \label{chipois}
\end{equation}
asymptotically obeys a chi-square distribution
(Baker and Cousins 1984).
It is easy to prove that the minimization of $\chi^2$ is equivalent to
maximization of the likelihood function,
so that the $\chi^2$ statistic may be useful both for
estimating the event characteristics and for goodness-of-fit testing.

The \textsc{CERN-MINUIT} package (James 1994)
was used to minimize (\ref{chipois}). The starting value for the i-th
coordinate was based on the charge asymmetries measured by initially
grouping the PMTs into only 6 ``superpatches'', referred to the detector
frame axes; it was defined according to:
\begin{equation}
  x_{i0} =\frac{\sqrt{Q^i_+}-\sqrt{Q^i_-}}{\sqrt{Q^i_+}+\sqrt{Q^i_-}}
D^i,
  \,\,\,i=1,2,3,
  \label{start}
\end{equation}
where the indices $+-$ refer to the opposite superpatches of the i-th
axis and
$D^i$ is the half size of the detector along that axis. Once the
$x_{i0}$
corresponding to the starting position is known, the starting energy
value is
obtained from (\ref{nthe}) after replacing $\vec{x}$ with $\vec{x}_0$
and
${\overline{N}_j}$ with ${N_j}$. Examples of the results obtained by
this procedure are shown in Fig. \ref{fig:chooz_Cf}.

\subsection{Results and systematics}

A summary of systematic errors for both \textsc{Chooz} and 
\textsc{Palo Verde}  is given in Table~\ref{tab:syst}.
The systematic
error given for \textsc{Chooz} should probably be considered 
as some sort of ultimate limit for 
reactor-based oscillation experiments, at least 
when only one detector is present.   Indeed
the intrinsically high efficiency ($\simeq 70\%$) 
of the homogeneous detector, together with the 
unique opportunity of studying the zero power case, 
are important advantages (for comparison
the efficiency of the larger but segmented 
\textsc{Palo Verde} detector is $\simeq 11\%$).

The (energy averaged) ratio between \nuebar detected 
and expected was found to be
\begin{equation} 
R= 1.01\pm2.8\%\rm(stat) \pm 2.7\%(syst) \;\;\;\;\;\;\; \textsc{Chooz}
\end{equation}
and 
\begin{equation} 
~~R= 1.01\pm2.4\%\rm(stat) \pm 5.3\%(syst) \;\;\;\;\;\;\; \textsc{Palo Verde}
\end{equation}
in both cases consistent with unity.

\begin{table}[h]
\begin{center}
\caption{ Origin and magnitude of systematic errors in 
\textsc{Palo Verde} and 
\textsc{Chooz}.
Note that the two experiments offer different 
breakdowns of their systematics.  For simplicity
we do not show the systematics for the \textsc{Palo Verde} 
ON-OFF analysis. The \textsc{Palo Verde} results
are from the analysis of the full data set (Boehm  {\it et al.} 2001).}
\medskip
\label{tab:syst}
\begin{tabular}{|lcccc|}
Systematic                       & \multicolumn{2}{c}{\textsc{Chooz} (\%)} 
& \multicolumn{2}{c}{\textsc{Palo Verde} (\%)}	\\ 
\hline
$\sigma (\rm{\bar\nu_e} + p \rightarrow n + e^+)$ &  1.9  &       &    -   &        \\
Number of p in target                             &  0.8  &       &    -   &        \\
$W_{th}$                                            &  0.7  &       &    -   &        \\
Energy absorbed per fission                       &  0.6  &       &    -   &        \\
Total rate prediction                             &       &  2.3  &        &   2.1  \\
e$^+$ trigger eff.                                   &   -   &       &   2.0  &        \\
n trigger eff.                                       &   -   &       &   2.1  &        \\
\nuebar selection cuts                            &   -   &       &   2.1  &        \\
$(1 - \epsilon_1)B_{\rm pn}$ estimate                             &   -   &       &   3.3  &        \\
Total \nuebar efficiency                          &       &  1.5  &        &   4.9  \\
\hline
Total                                             &       &  2.7  &        &   5.3  \\
\end{tabular}
\end{center}
\end{table}

\begin{figure}[h]
\begin{center}
\epsfysize=4.3in \epsfbox{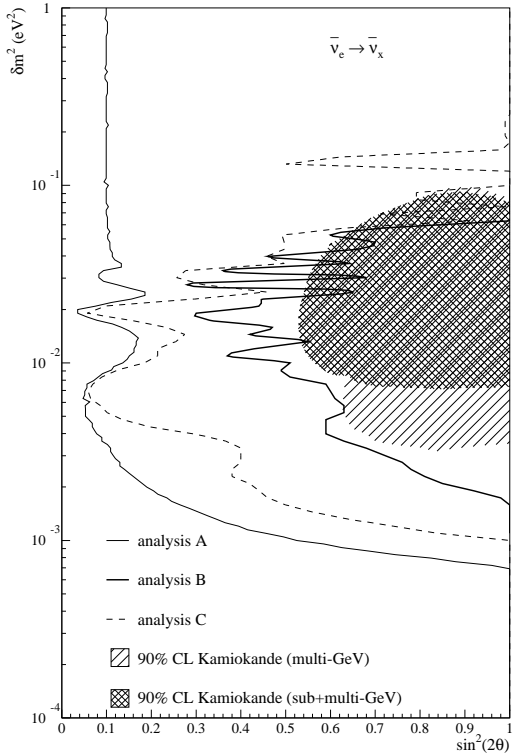}
\caption{ Limits on mass difference and mixing angle from 
\textsc{Chooz} (90\%~CL)
obtained with the unified approach
(Feldman and Cousins 1998).
Analysis A refers to the curve obtained
by a fit to the background-subtracted spectrum
in which both shape and normalization are
used.   Analysis C uses only the shape of the spectrum.
Finally analysis B uses the difference
of baselines between the two reactors ($\Delta L = 116.7$~m).
While in this last case
most systematics cancel, statistical errors
are larger and the $\Delta m^2$ sensitivity
rather poor due to the short baseline difference.
The Kamiokande $\rm \nu_e - \nu_{\mu}$ atmospheric neutrino
result is also shown.}
\label{fig:chooz_excl}
\end{center}
\end{figure}

\begin{figure}[h]
\begin{center}
\epsfysize=3.3in \epsfbox{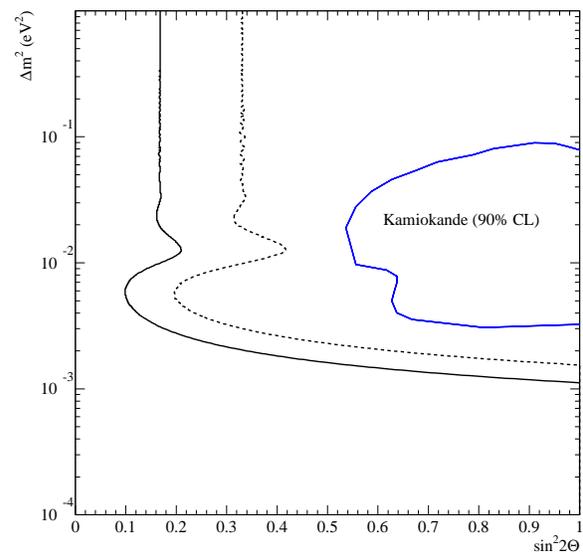}
\caption{ Limits on mass difference and mixing angle from 
\textsc{Palo Verde} at 90\%~CL (Boehm  {\it et al.} 2001).
The full curve  is obtained with
the {\it swap} background subtraction method
described in the text, while the dashed
is obtained  using the reactor
power changes to estimate and subtract the
background.
The Kamiokande $\rm \nu_{\mu} - \nu_e$ atmospheric neutrino
result is also shown for illustration.}
\label{fig:pv_excl}
\end{center}
\end{figure}

Both experiments were able to exclude \nuebar - $\bar\nu_{\rm x}$ oscillations
as dominating  for the atmospheric neutrino anomaly.    
This is evident from the exclusion
contours obtained  using 
the unified approach (Feldman and Cousins 1998)
and shown in 
Figure~\ref{fig:chooz_excl} for \textsc{Chooz} and 
Figure~\ref{fig:pv_excl} for \textsc{Palo Verde}.

\subsection{Are smaller mixing angles within experimental reach ?}

The current data on neutrino oscillations suggests 
the need to include at least 
three neutrino flavors when studying results from experiments.
As discussed in the Introduction, 
the most general approach would involve five unknown parameters, three
mixing angles and two independent mass differences. However, an intermediate 
approach consists of a simple generalization 
of the two flavor scenario, assuming 
that $m_3^2 \gg m_1^2, m_2^2$ (i.e. 
$\Delta m_{13}^2 = \Delta m_{23}^2 = \Delta m^2$, 
while $ \Delta m_{12}^2 \simeq 0$). 
This scenario is obviously compatible with the evidence based
on the atmospheric neutrino anomaly ($\Delta m^2 \sim 3 \times 10^{-3}$ eV$^2$)
and the solar neutrino deficit ($\Delta m^2 < 10^{-4}$ eV$^2$).
In such a case the mixing angle $\theta_{12}$ 
becomes irrelevant and one is left with only three unknown quantities:
$ \Delta m^2, \theta_{13}, {\rm and} ~\theta_{23}$. 
With this parameterization, and 
assuming that $\nu$ behave like $\bar\nu$, 
the $\bar{\nu}_e$ disappearance is governed by
\begin{equation}
P(\bar{\nu}_e \rightarrow \bar{\nu}_x) = 
\sin^2 2\theta_{13} \sin^2 \frac{\Delta m^2 L}{4E_{\nu}} ~,
\end{equation}
while $\nu_{\mu} \rightarrow \nu_{\tau}$ oscillations, 
in this scenario responsible 
for the atmospheric neutrino results, are described by
\begin{equation}
P(\nu_{\mu} \rightarrow \nu_{\tau}) = 
\cos^4 \theta_{13} \sin^2 2\theta_{23}  
\sin^2 \frac{\Delta m^2 L}{4E_{\nu}} ~.
\end{equation} 

\begin{figure}[htb!!!]
\begin{center}
\epsfysize=2.8in \epsfbox{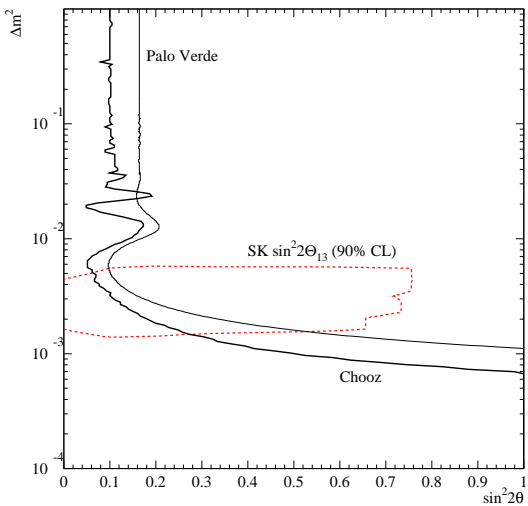}

\caption{ Exclusion plot showing the $allowed$ region of $\theta_{13}$
and $\Delta m^2$ based on the Super-Kamiokande preliminary analysis
(the region $inside$ the dotted curve). The region
$excluded$ by the neutrino reactor experiments
are to the right of the corresponding thick and thin continuous curves.}
\label{fig:SK}
\end{center}
\end{figure}

An analysis of the atmospheric neutrino data based on these 
assumptions has been performed (Okamura 1999)
and its results are shown in 
Figure~\ref{fig:SK} for the $\nu_e$ disappearance channel. One can 
see that, while the relevant region of the mass difference $\Delta m^2$
is determined by the atmospheric neutrino data, the mixing angle 
$\theta_{13}$ is not constrained very much. 
Here the reactor-based neutrino oscillation experiments play a decisive role.

The determination of the angle $\theta_{13}$ has  obvious importance
not only for the structure of the lepton mixing matrix $U$ but
for the observability of $CP$ violation in the lepton sector,
as stressed in the Introduction. If  $\theta_{13}$ vanishes,
or is very small, no $CP$ violation effects are observable
in the lepton sector. 
Moreover, for the vanishing $\theta_{13}$ and with three neutrinos
only, the lepton mixing is radically simplified.
The electron neutrino is then simply
\begin{equation}
\nu_e = \cos\theta_{12}\nu_1 + \sin\theta_{12}\nu_2~,
\end{equation}
while the $\nu_{\mu}$ and $\nu_{\tau}$ neutrinos become
superpositions of $\nu_3$ and the corresponding orthogonal
combination of  $\nu_1$ and  $\nu_2$. 
It is therefore
interesting to ask whether \rb experiments can be extended to address regions
of even smaller mixing parameter $\sin^2{2\theta_{13}}$. 

A simple inspection of Table~\ref{tab:syst} shows that, using the
\textsc{Chooz} systematics, if all flux and cross-sections related errors could
be set to zero one would be left with an error of $\simeq 1.5$\%.  Hence, assuming 
a detector large enough to produce negligible statistical error
the total error would shrink from the present 3.9\% to 1.5\%.   

This scenario is considered by Mikaelyan (2000)
that proposes to use
an underground reactor at Krasnoyarsk 
in Russia as a source and two identical detectors
placed at distances of $\simeq 1100$~m and $\simeq 250$~m.   The interesting
feature of the Krasnoyarsk site is that there are substantial facilities
available underground, with an overburden of $\sim 600$~m.w.e., twice
the depth of \textsc{Chooz}.   
Indeed it might even be conceivable to locate the 
detectors on rail-cars and periodically switch their position to further
reduce some of the systematics related to detector efficiency.    
The proposal discusses
the use of 50~tons of Gd-loaded scintillator for each of the two 
identical homogeneous 
detectors, so that the far detector would collect 50~events/day 
(the thermal power
of the reactor is in this case lower than at 
\textsc{Chooz} or \textsc{Palo Verde}).  
The background is
estimated by Mikaelyan (2000)
to be of 5~events/day or less.    

This proposal estimates that such an experiment could reach a 
sensitivity in mixing strength of better than 0.02 in the $\Delta m^2$
region relevant for atmospheric neutrinos.    
While the idea looks certainly interesting, it would be useful to explore
how practical it is in general to push the errors of the absolute \nuebar
flux to the 1\% domain, even with the measurements considered here.
Note also that the 
Krasnoyarsk reactors, according to the 
Gore-Chernomyrdin\footnote{``US-Russian Plutonium Production
Agreement is Signed'' Statement by the White House office of
the Vice President, 23 Sept. 1997.}
agreement, are supposed to be shutdown for re-coring in not very distant future.

\section{Exploring the solar $\nu$ anomaly on earth: \textsc{Kamland}}

While historically solar neutrinos provided the first hint for oscillations,
there is a consensus today that the strongest evidence for oscillation is  the
atmospheric neutrino anomaly.     Indeed, the zenith angle dependence of the anomaly has 
substantially helped to eliminate explanations
not based on some property of neutrinos themselves and the 
advent of K2K, to be followed soon by the \textsc{Minos} and CERN to Gran Sasso 
programs (Wojcicki 2001a, 2001b), are bringing the study of oscillations in this regime 
to a laboratory activity with both source and detector well under control.

In the case of solar neutrinos none of the effects that would be generally
considered ``smoking guns'' for oscillations has yet clearly emerged from the data
and
their exploration ``in a laboratory setting'' is made particularly challenging by
the huge $L/E_{\nu}$ required.    It is probably a safe prediction that it will 
take a very long time before an \ab experiment will be able to tackle the solar
neutrino problem!   However, the very low energy of reactor neutrinos make a \rb
oscillation experiment able to reach the Large Mixing Angle (LMA MSW)  
solution possible - albeit rather challenging.    While 
the analysis of current and future solar neutrino experiments 
presumably will help to decide which of the

\onecolumn

\begin{figure*}[ptb!!!]
\begin{center}
\epsfysize=3.1in \epsfbox{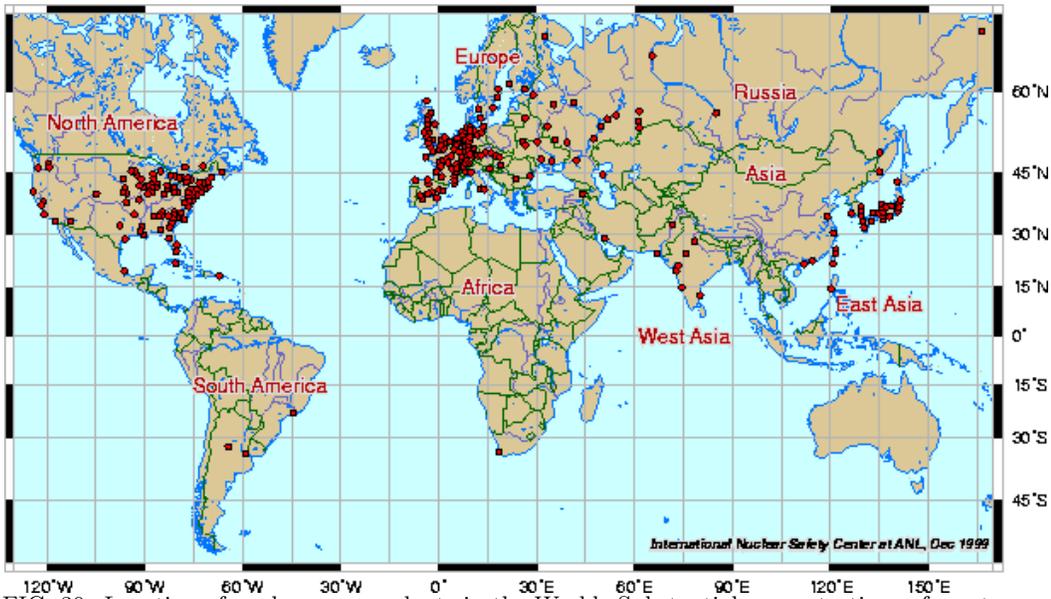}
\caption{ Location of nuclear power plants in the World.   Substantial
concentrations of reactors are in Europe, East of the US and Japan.
(Note that the map, see  http://www.insc.anl.gov/, 
contains few plants that were either planned
but never built or are no longer operational.)}
\label{fig:world_nukes}
\end{center}
\end{figure*}
\begin{figure*}[ptb!!!]
\begin{center}
\epsfysize=4.2in \epsfbox{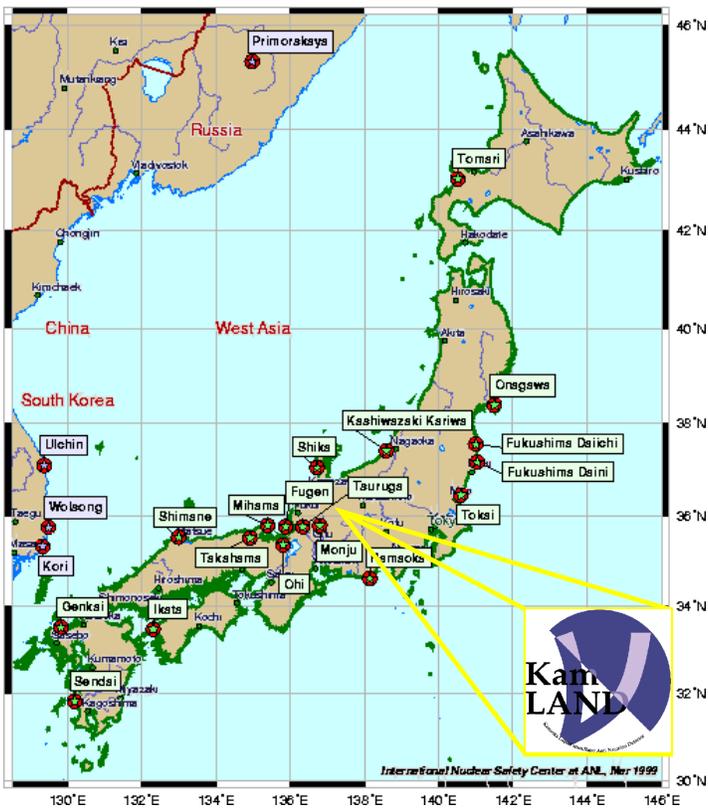}
\caption{ Location of large nuclear power plants in Japan, Korea and
Far East Russia. (See the comment in the caption of Fig. 
\protect\ref{fig:world_nukes}.)}
\label{fig:japan_nukes}
\end{center}
\end{figure*}

\twocolumn[\hsize\textwidth\columnwidth\hsize\csname
@twocolumnfalse\endcsname]

\noindent 
possible solutions is the right one,
we find the 
chance to study solar neutrino oscillations ``in the lab'' extremely 
compelling.  Note that, unlike in the case of atmospheric neutrinos, where
it turned-out that electron neutrinos are not involved in the dominant mixing,
the solar neutrino problem, if due to oscillations,  obviously involves  
$\nu_e$ disappearance.   So, unless
\nuebar behave drastically differently from $\nu_{\rm e}$ (which would be a
worthwhile discovery anyway, signaling the
breakdown of CPT symmetry) a reactor experiment is an exact replica of the 
astrophysical experiment, only built on earth.

\subsection{Nuclear reactors in Japan}

\begin{figure}[htb!!!]
\begin{center}
\epsfysize=3.2in \epsfbox{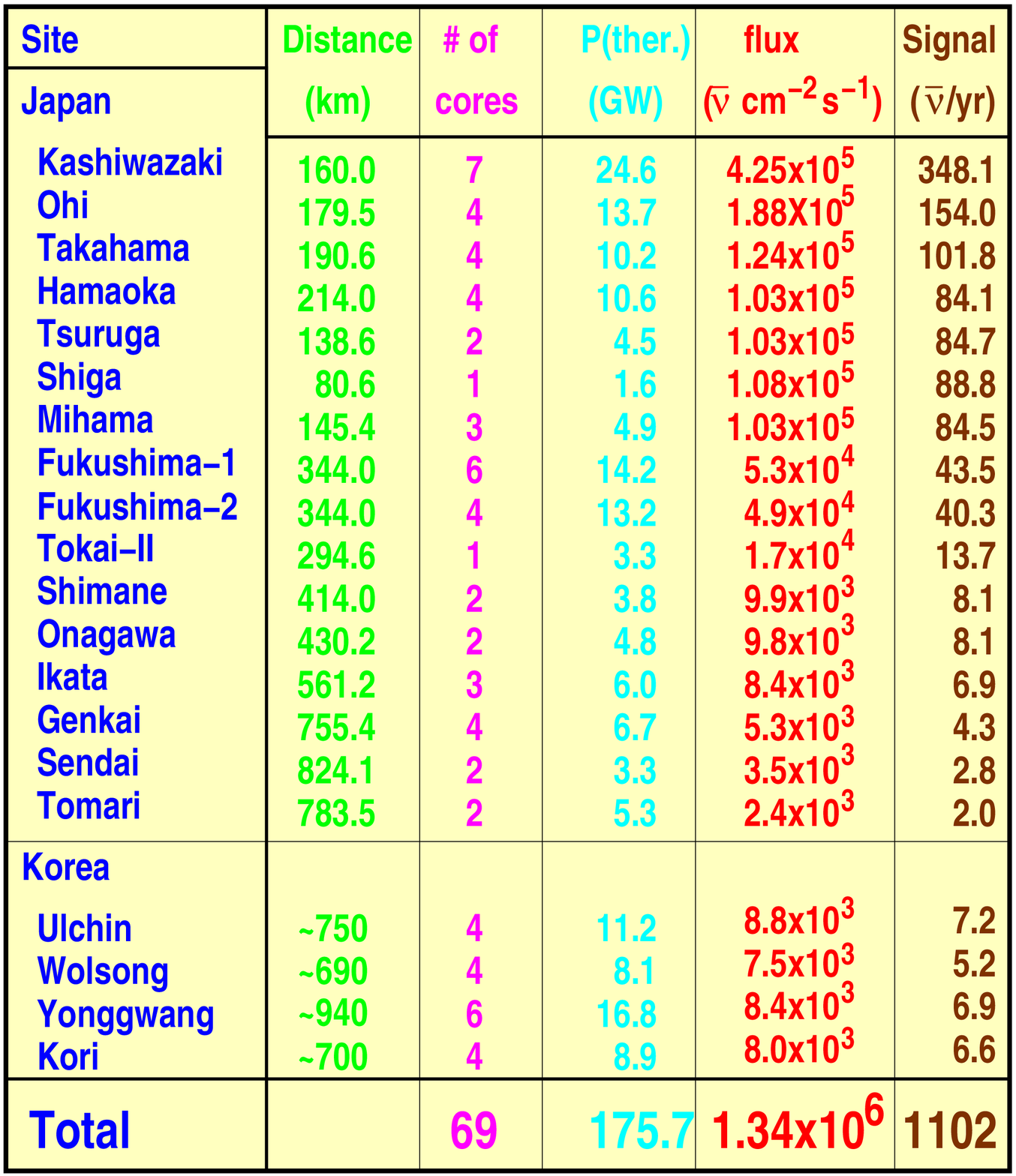}
\medskip
\caption{ List of relevant parameters for power reactors in Japan and
South Korea.}
\label{fig:reactor_table}
\end{center}
\end{figure}

The ``easier'' solution of the solar neutrino problem (MSW LMA) is shown in 
Figure~\ref{fig:sensi}.   In order to completely explore such solution one needs a 
$\Delta m^2$ sensitivity of at least $10^{-5}$~eV$^2$ at large mixing angle.
As now customary we refer to Figure~\ref{fig:rate_mass} as a first step in designing
our experiment: we see that a $\approx 100$~km baseline is needed and this drives the 
power$\times$fiducial-mass product in the $10^8$ MW$_{th}\times$tons range.   Clearly
a large detector has to be used in conjunction with very many nuclear reactors.
A cursory look at the placement of nuclear power plants on the earth, 
Figure~\ref{fig:world_nukes}, reveals that such an experiment could only be placed
in Europe, East of the United States, or Japan.    
There are 16 commercial nuclear power plants in Japan, their location being shown in
Figure~\ref{fig:japan_nukes}.   They supply about 1/3 (or 130 $\rm GW_{th}$) 
of the total electric power in the country.
At the Kamioka site there is an anti-neutrino 
flux of $\simeq 4\times 10^6$cm$^{-2}$s$^{-1}$
(or $\simeq 1.3\times 10^6$cm$^{-2}$s$^{-1}$ 
for $E_{\bar\nu} > 1.8$MeV) from these reactors.
80\% of such flux derives from reactors at a distance between 140~km and 210~km, so 
that there is a limited range of baselines.  
The breakdown of this data by power plant (several plants have on site more than one
reactor) is given in Figure~\ref{fig:reactor_table}.   We note that some 2\% of the flux
derives from power plants in South Korea 
(the Primorskaya plant in Russia is only planned) 
that will have to be included (albeit only as a 
crude estimate) to provide an exact flux prediction. 

\begin{figure}[h]
\begin{center}
\epsfysize=2.3in \epsfbox{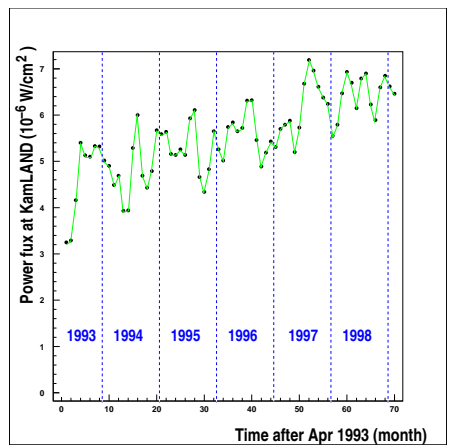}
\medskip
\caption{ Power-flux level at Kamioka from Japanese reactors as function
of time. Low power periods in the Fall and Spring seasons are alternated with peaks of
high power in the Summer and Winter.}
\label{fig:power_excursion}
\end{center}
\end{figure}

While the Figure~\ref{fig:reactor_table} 
assumes the nominal power for each of the cores, an average over one
year, taking into account scheduled and unscheduled down times, gives an expected
non-oscillation rate of $\simeq$750~kton$^{-1}$year$^{-1}$ for a C$_n$H$_{2n+2}$ target.
Although the signal is provided by a very large number of cores it turns out that a 
modulation of the \nuebar flux is expected at 
\textsc{Kamland} (Alivisatos {\it et al.} 1998) thanks to 
the refueling and maintenance schedule of nuclear power plants in Japan.   
Such shutdowns,
in fact, are concentrated in the Fall and Spring when the power demand is lowest, as
illustrated in Figure~\ref{fig:power_excursion}.  Hence, from the point 
of view of the
tools available to study backgrounds, 
\textsc{Kamland} is in a situation very similar to that of
\textsc{Palo Verde}, with 2 dips in the flux from full to 
$\approx 2/3$ expected every year.


It is interesting to remark that other artificial sources of low energy \nuebar
are not a serious background in \textsc{Kamland}. The largest effect would be produced
by a large nuclear powered vessel stationed, while running its reactors
at full power, in the Toyama bay, 50 - 100 km from the detector. In these
circumstances the excess signal in  \textsc{Kamland} would amount to $\sim$10\%
(Detwiler 2000). It is clear that it is extremely unlikely that
such conditions will occur for any significant period of time.

\subsection{Detector design}

\begin{figure*}[htbp]
\begin{center}
\epsfysize=3.6in \epsfbox{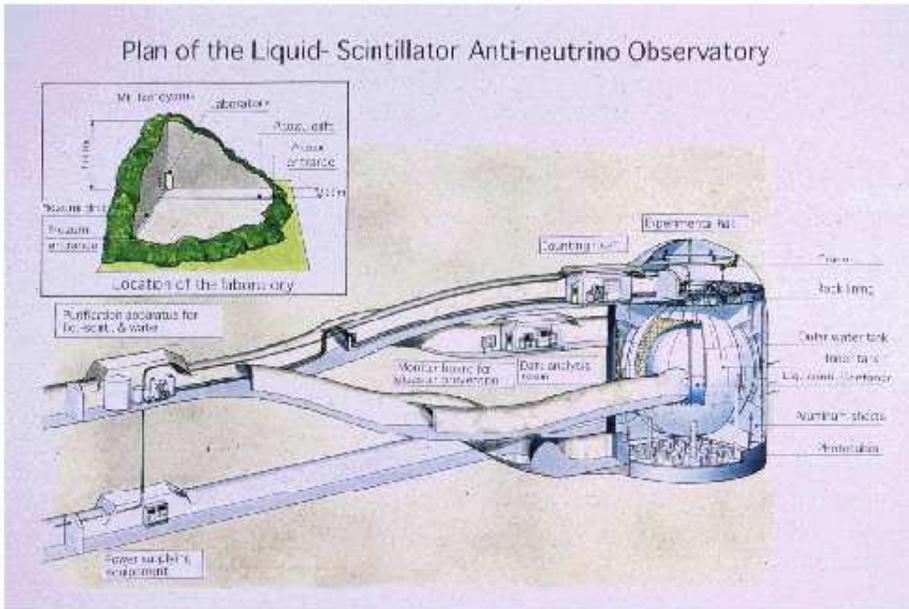}
\caption{ Partial view of the system of tunnels inside Mount
Ikenoyama with the locations of \textsc{Kamland} and its main services.}
\label{fig:ikenoyama}
\end{center}
\end{figure*}

The \textsc{Kamland} detector is housed in the cavity built for the Kamiokande detector under the 
summit of Mt. Ikenoyama in the Japanese Alps, about 50~km east of the town of Toyama.
The layout of the laboratory is 
shown in Figure~\ref{fig:ikenoyama}.
The rock overburden is more than 1,000~m in 
any direction with an average rock density 
of 2.7~g/cm$^3$.  The site is at 500~m distance from SuperKamiokande.

\begin{figure*}[htbp]
\begin{center}
\epsfysize=4.4in \epsfbox{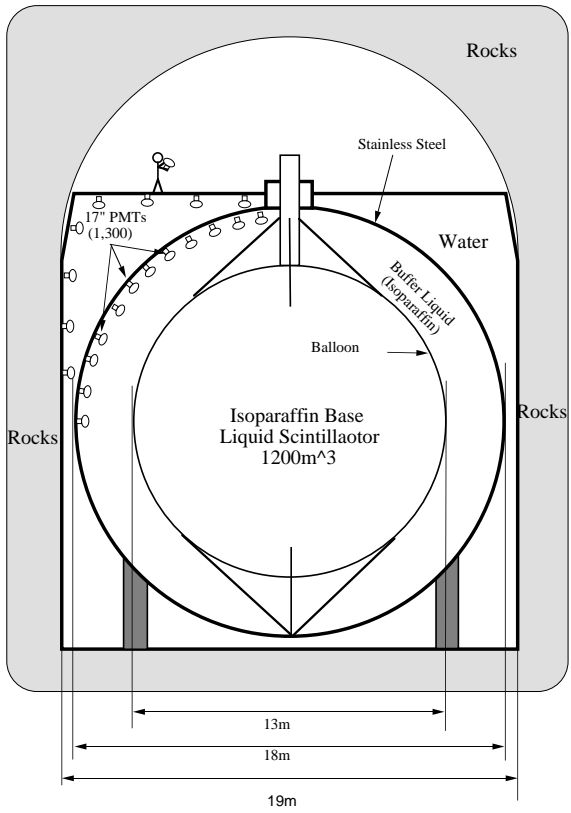}
\caption{ Schematic cross-section of the \textsc{Kamland} detector.}
\label{fig:Kamland_det_setup}
\end{center}
\end{figure*}

A cutout view of the \textsc{Kamland} detector is shown in 
Figure~\ref{fig:Kamland_det_setup}.  
The fiducial volume consists of a sphere containing 1000 tons of liquid scintillator.  
The scintillator container is a thin plastic-walled balloon of 6.5~m radius that is 
not supposed to take the weight of the scintillator but only to isolate it from an 
outer 2.5~m thick layer of non-scintillating, radiation shielding, fluid. 
The balloon is also designed to be impermeable to radon that mainly originates from 
Th and U contaminations inside the PMT's glass.   The buffer fluid and the liquid 
scintillator are contained and mechanically supported by a 
stainless steel spherical 
vessel that also provides the mechanical structure where the photomultipliers for 
the fiducial volume are mounted.    
The sphere is solidly anchored inside the cylindrical rock cavity and
the space between them is filled with water and used as a veto \v{C}erenkov counter.
The scintillator, based on mineral oil and pseudocumene, is designed to achieve
sufficient light yield and n-$\gamma$ discrimination by pulse-shape analysis,
yet complying with rather strict flammability requirements from the Kamioka mine.
Given the cost and stability issues for a detector of the size of \textsc{Kamland}, it
was chosen not to Gd-load the scintillator.   As it will be discussed later, simulations
indicate that sufficient signal-to-noise ratio 
will be achieved with unloaded scintillator.
Events will be localized inside the fiducial volume using the light 
intensity and propagation delays
to the different photomultipliers so that large area, fast tubes are required.   
While the veto counter will be read-out using 20-inch photomultipliers dismounted from 
the Kamiokande detector, new, faster, tubes with 17-inch active photocathode have been 
developed for \textsc{Kamland} in order to allow for proper vertex reconstruction from timing.   
Such tubes have an average transit-time-spread of 
$\simeq 3$~ns (to be compared to $\simeq 5$~ns for the  Kamiokande/SuperKamiokande 
tubes).   The central detector has a 30\% photocathode coverage obtained using about 
1280 seventeen-inch tubes complemented, 
for energy measurements, by 642 twenty-inch Kamiokande 
tubes.   A spherical shell of acrylic panels (not shown in 
Figure~\ref{fig:Kamland_det_setup}) is mounted at a radius immediately inside the 
position of the PMT's and is used as 
primary barrier against radon migration into 
the active scintillator.
A cylindrical stainless steel chimney of 3~m diameter protrudes from the top of 
the sphere to permit access to the central detector during installation. Buffer fluid 
and scintillator lines as well as calibration ports are mounted in the chimney 
along with all the electrical cabling.

The readout of \textsc{Kamland} is designed to provide waveform analysis information for each of
the PMT's in the detector with essentially no dead-time for several consecutive events.
This allows for clean event reconstruction 
and enables the off-line study of the pre-history
of interesting events.   For example multiple neutron events, described above as the most 
dangerous background at \textsc{Palo Verde}, 
will be fully reconstructed by \textsc{Kamland}.   Similarly
cosmogenic activation giving short half-life nuclei will be clearly recorded.
Deep digital buffering will allow the detector to sustain substantial burst of events
like expected from supernovae.

\begin{figure*}[tbp!!!!!!!!!!!!!!!!!!!!!!!!!!]
\begin{center}
\epsfysize=4.2in \epsfbox{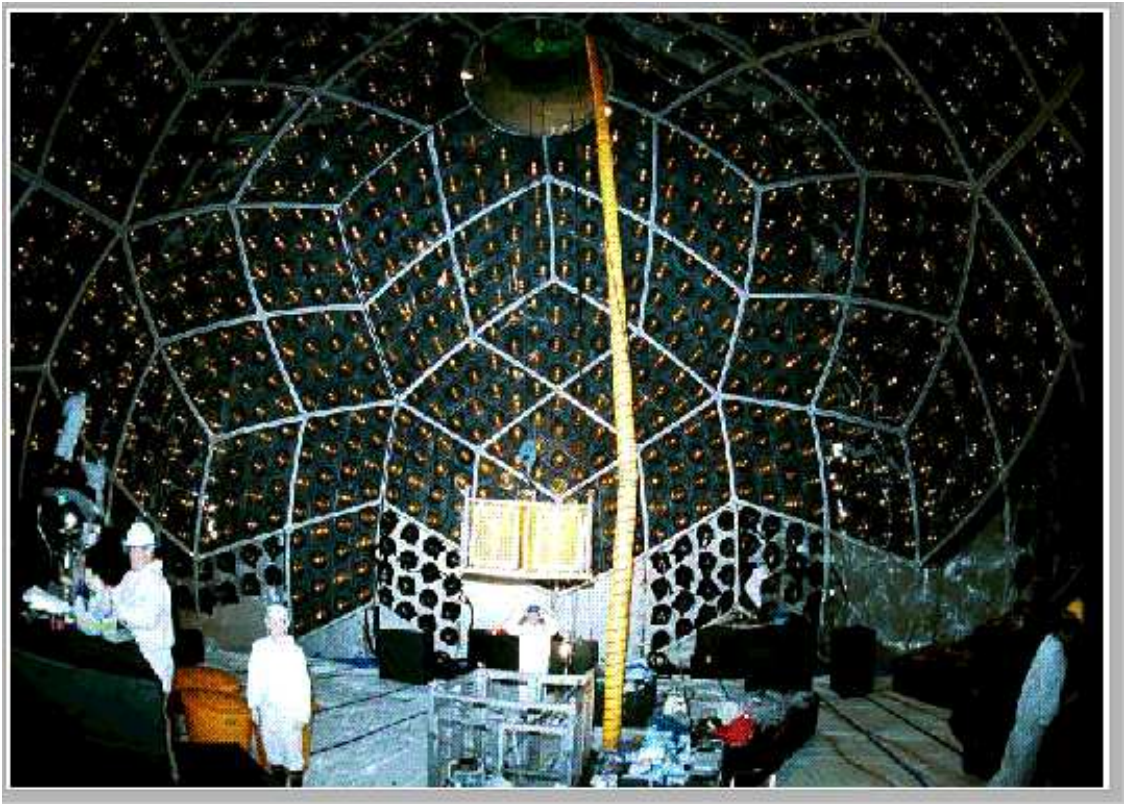}
\medskip
\caption{ View of the internal volume of the 
\textsc{Kamland} sphere during the
central detector installation.    A modular styrofoam raft is used as a platform
for workers. The installation begun from the top of the sphere and moved down,
as the water level in the sphere was reduced.   PMTs, black shades, acrylic plates,
monitoring LEDs and cables were mounted in place for each level before lowering the
water.}
\label{fig:PMT_inst}
\end{center}
\end{figure*}

In Figure~\ref{fig:PMT_inst} we show a phase of the central detector PMT installation
that was concluded in September 2000. Scintillator
filling started in Spring 2001 and data taking
at  \textsc{Kamland} is  scheduled to begin before the end of  2001.

\subsection{Expected performance}

Similarly to previous experiments, both random hits from natural radioactivity 
and correlated events from neutron production in cosmic-ray-muon spallation 
and capture, contribute to the background to reactor \nuebar in \textsc{Kamland}.
The results of Monte Carlo full detector simulation using the measured
Kamioka cosmic ray flux and the activities of various components as sampled
during construction are given in Table~\ref{tab:sumb}.   
For the purpose of this background estimate  U and Th contaminations
in the scintillator of $10^{-14}$~g/g has been assumed.
Such purity level has been already achieved in samples
of the \textsc{Kamland} scintillator.
Monte Carlo studies have shown that cosmogenic activation gives 
negligible contribution to the background for doubles.    A discussion of 
backgrounds to single signatures, not considered here, can be found
in Alivisatos {\it et al.} (1998).    

\begin{table}[htb]
\begin{center}
\caption{ Summary of background rates 
in \textsc{Kamland} for the \nuebar signature.
A signal-to-noise ratio of about 10/1 is expected for reactor \nuebar.
Adapted from Alivisatos {\it et al.} (1998).}
\begin{tabular}{|l|c|}
Background source                          & Rate (day$^{-1}$) \\
\hline
Cosmic muons induces neutrons              &        0.1        \\
Natural radioactivity (random coincidence) &        0.15       \\
Natural radioactivity (correlated)         &        0.005      \\
\hline
Total predicted background                 &        0.25       \\
Reactor \nuebar signal (no oscillation)    &        2          \\
\end{tabular} 
\label{tab:sumb}
\end{center} 
\end{table}

\onecolumn

\twocolumn[\hsize\textwidth\columnwidth\hsize\csname
@twocolumnfalse\endcsname]

\begin{figure}[h]
\begin{center}
\epsfysize=3.2in \epsfbox{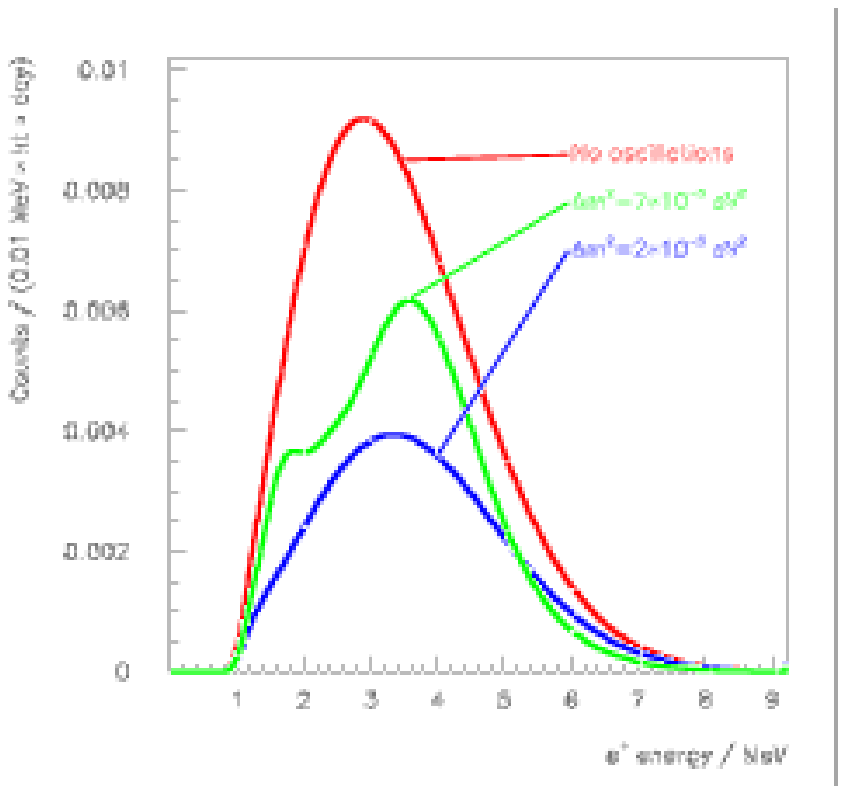}
\caption{ Positron energy spectra expected at \textsc{Kamland} for no oscillations
and oscillations with 
indicated parameters $\Delta m^2$ and
$\sin^2 2\theta = 0.75$ in the MSW LMA  solar neutrino solution.}
\label{fig:kl_spectra}
\end{center}
\end{figure}

In Figure~\ref{fig:kl_spectra} we show the predicted energy spectra for reactor 
neutrinos at \textsc{Kamland} for no oscillations and different oscillation parameters 
consistent with the LMA MSW solar neutrino solution.
We can use one of such curves and add to it fluctuations consistent with a 10/1
signal-to-noise ratio and three years of data to investigate the sensitivity of the
experiment. Assuming that oscillations with $\Delta m^2 = 2\times 10^{-5}$~eV$^2$ 
and $\sin^2 {2\theta} = 0.75$ are indeed the cause of the solar neutrino anomaly,
we obtain the measurement of the oscillation parameters shown in 
Figure~\ref{fig:evidence}.    On the other hand, no evidence for oscillation after
three years of data would result in the exclusion curve shown in Figure~\ref{fig:sensi}
and would rule-out the LMA MSW solution to the solar neutrino problem.

\subsection{Other physics with a very large 
low-energy \nuebar detector}

\textsc{Kamland} will be the largest detector 
specifically optimized to detect low-energy \nuebar
with good efficiency and low background.   
This opens a number of interesting opportunities
beyond the measurement of oscillations from reactors.     
In addition, such a large detector
with a low energy threshold can be used to directly measure 
neutrinos from the sun, 
assuming that backgrounds can be
sufficiently reduced and understood to enable the
detection
of single energy deposits. Of particular importance is
the $^7$Be line that is below the threshold of the water \v{C}erenkov detectors.   
The presence of large amounts of carbon in \textsc{Kamland}'s scintillator 
opens the possibility of detailed flavor
studies in neutrinos coming from supernovae.  
Finally \textsc{Kamland} represents such a large step in 
size and backgrounds relative to the previous detectors that one should be ready for the 
possibility that it will discover completely new and un-expected phenomena in physics or 
astrophysics.

\begin{figure}[htbp]
\begin{center}
\epsfysize=2.8in \epsfbox{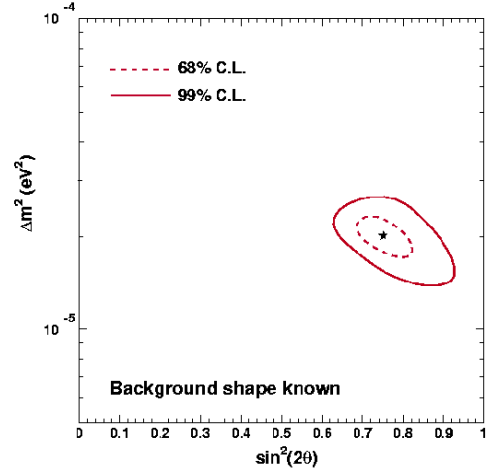}
\caption{ Simulated measurement of neutrino oscillation parameters from
three years of data at \textsc{Kamland} assuming  $\Delta m^2 = 2\times 10^{-5}$~eV$^2$
and $\sin^2 {2\theta} = 0.75$.  A signal-to-noise ratio of 10/1 was assumed (see text).}
\label{fig:evidence}
\end{center}
\end{figure}

Here  we will only mention  the topic of terrestrial \nuebar
that is somewhat unusual and directly relates to the experiment's 
ability to detect \nuebar.   The reader interested in the direct detection of solar 
neutrinos or neutrinos from supernovae is referred to the \textsc{Kamland} design 
report (Alivisatos {\it et al.} 1998).   
A description of the new and un-expected phenomena 
mentioned above will be hopefully provided at a later stage.

Although the study of terrestrial anti-neutrinos 
was proposed as early as 1966 (Eders 1966)
practical difficulties, 
due to the very small cross-sections and very low energies involved,
have made this physics impractical until now.   
\textsc{Kamland} has the ability to detect energy
depositions of the order of 1~MeV in a unprecedented amount of liquid scintillator
and is therefore ideally suited for this study.    It is important to realize that low
energy \nuebar are easily detected with very low background in \textsc{Kamland} thanks to
their very specific signature.

The cooling rate of our planet and its contents of heavy elements are central issues
in the earth sciences.
The earth radiates about 40 TW of heat from its surface.  
About 40\% of this energy (or 16 TW)
is believed to have radiogenic origin with 90\% of 
it deriving from decays of $^{238}$U and 
$^{232}$Th.   Radiogenic heat is therefore an essential component 
of the present Earth dynamics. 
As discussed by several authors (Eders 1966, Marx 1969, Marx and Lux 1970,
Avilez {\it et al.} 1981, Krauss 1984) 
the concentration of these isotopes
can be mapped, at planetary scale, by direct detection of \nuebar deriving
from the $\beta$-decay processes.   Since neutrinos have a mean free path many orders
of magnitude larger that the size of the Earth, the neutrino field is analogous 
to a gravitational field, 
where the sources are represented 
by radioactive density (as opposed to mass density).

\begin{figure}[htbp!!!!!!!!!!!!!!]
\begin{center}
\epsfysize=2.5in \epsfbox{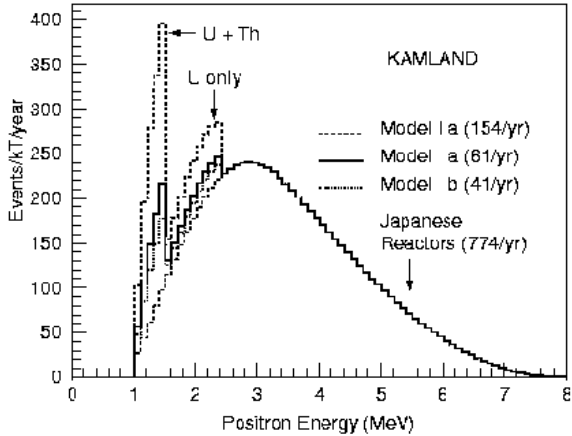}
\caption{ Energy spectrum from terrestrial anti-neutrinos compared with reactor signal as
expected in \textsc{Kamland}.  Three different geophysical models 
are shown for the terrestrial
anti-neutrinos and no oscillations are assumed for all the spectra shown.
Adapted from Raghavan {\it et al.} (1998).}
\label{fig:k_terr_neu}
\end{center}
\end{figure}

Since the maximum energy carried by terrestrial neutrinos
is (Krauss 1984) 
3.27~MeV and the capture threshold is 1.8~MeV, the maximum in the energy 
spectrum detected in the prompt part of the 
events will be 2.49~MeV (including the 1.02~MeV from positron annihilations).      
For energies above threshold only the Thorium and Uranium 
decay chains give a detectable amount of events. 
$^{234}$Pa from the U chain and $^{228}$Ac and
$^{212}$Bi of the Th chain have similar endpoints 
(respectively 2.29~MeV, 2.08~MeV and 2.25~MeV)
while $^{214}$Bi from the U chain has an endpoint of 3.27~MeV.  Therefore
the energy spectrum observed for the prompt part of the event should
have a characteristic 
double-hump structure shown in Figure~\ref{fig:k_terr_neu}. This will also allow the 
measurement of the U/Th ratio.   Anti-neutrinos from 
nuclear reactors give, as described above, a similar signature, but their energy is
substantially higher and, as shown in the Figure, they can be easily separated from the
terrestrial anti-neutrinos.  
Indeed, a repetition of the analysis for the reactor neutrinos
discussed above using only positron energies above 2.7 MeV gives
an oscillation sensitivity very similar to the one
presented in Figure~\ref{fig:evidence}.

The two lower spectra (Ia and Ib) superimposed in Figure~\ref{fig:k_terr_neu} for the 
terrestrial anti-neutrino component correspond to two different possible geophysical 
models with different heavy elements concentration in the oceanic and continental
crusts (Raghavan {\it et al.} 1998).   
The highest curve (IIa) is given as a reference and shows 
what the spectrum would be in the extreme case where the entire 40 TW of heat escaping 
from the Earth's interior was generated by the Th and U decay chains.

In one year of data-taking, model Ia would give an integral of 61 events while model Ib 
would give only 41 events, and a differentiation between the two at 3$\sigma$ level
could be obtained in five years of data-taking, taking into account the fluctuations of
the background due to the reactor neutrinos.

\section{Conclusions}

The use of nuclear reactors to study neutrino properties has a long and glorious
history.   While the first experiments devoted to oscillation searches were 
motivated by the generic principle to ``look where the light is'', many of the
modern hints for neutrino oscillations point to parameters that match very well
the capabilities of reactor-based experiments.   At the same time the understanding
of the flux and spectrum of \nuebar from power reactors has reached substantial
sophistication.     The first two ``long baseline'' experiments, 
\textsc{Chooz} and  \textsc{Palo Verde}, 
have amply demonstrated the capability of this new breed of detectors, while
providing solid evidence that $\rm\nu_e - \nu_{\mu}$
is {\it not} the dominant channel
in the atmospheric neutrino oscillation.
\textsc{Kamland}, scheduled to begin data-taking shortly, will extend the reach 
for small mass-differences to unprecedented levels.   Its size and background will
move \rb experiments to a new dimension, with several new physics opportunities
in the essentially background-less detection of \nuebar from a number of natural
sources.

\section{Acknowledgment}

We would like to thank our colleagues from  \textsc{Chooz}, 
\textsc{Kamland} and
 \textsc{Palo Verde} for the many discussions and help received in
understanding the material discussed.    This work was supported
in part by the US DoE 
and by the Italian INFN.


\begin{thebibliography}{10}

\bibitem{bugeyd}  Achkar B., 1992, Thesis, (ILL Grenoble).

\bibitem{bugey} Achkar B. {\it et al.}, 1995, Nucl. Phys. {\bf B434}, 503.

\bibitem{bugey3} Achkar  B. {\it et al.}, 1996,  Phys. Lett. {\bf B374}, 243.

\bibitem{SNO} Ahmad Q. R. {\it et al.}, 2001, Phys. Rev. Lett., submitted.

\bibitem{rovno} Alfonin  A. I. {\it et al.}, 1998, JETP {\bf 67}, 213.

\bibitem{borexino1} Alimonti G. {\it et al.}, 1998, Astroparticle Physics {\bf 8}, 141.

\bibitem{borexino2} Alimonti G. {\it et al}., 2000, NIM  {\bf A440}, 360.

\bibitem{KL_design_report} Alivisatos P. {\it et al}. 1998, \textsc{Kamland}: {\it
a liquid scintillator  Anti-Neutrino Detector at the Kamioka site}, preprint
Stanford-HEP-98-03, Tohoku-RCNS-98-15.

\bibitem{MUNU} Amsler C. {\it et al.}, 1997, Nucl. Inst. Meth. {\bf A396},
115.

\bibitem{chooz_osc_1} Apollonio M. {\it et al.}, 1998, Phys. Lett. 
{\bf B420}, 397.

\bibitem{chooz_osc_2} Apollonio M. {\it et al.}, 1999, Phys. Lett. 
{\bf B466}, 415.

\bibitem{position}  Apollonio M. {\it et al.}, 2000, Phys. Rev. D {\bf 61}, 012001.

\bibitem{Karmen}
Armbruster B. {\it et al.}, 1998,  Phys. Rev. C{\bf 57}, 3414.

\bibitem{LSND}  Athanassopoulos C. {\it et al.}, 1995,
Phys. Rev. Lett. {\bf 75}, 2650.
\bibitem{} Athanassopoulos C. {\it et al.}, 1996, Phys. Rev. C{\bf  54},
2685.
\bibitem{} Athanassopoulos C. {\it et al.}, 1998, Phys. Rev. C{\bf  58}, 2489.

\bibitem{} Avilez C. {\it et al.}, 1981, Phys. Rev. D{\bf 23}, 1116.

\bibitem{Bahcall}  Bahcall J. N., 1989,  {\it Neutrino Astrophysics},
(Cambridge Univ. Press, Cambridge).


\bibitem{chooz_det} Baldini A. {\it et al.}, 1996,
Nucl. Instr. and Meth. {\bf A372},207.

\bibitem{Cousins} Baker S. and Cousins R., 1984, Nucl. Instr. Meth.
{\bf 221}, 437.

\bibitem{cp2}Barger V., Whisnant K., and Phillips R. J. N., 1980,
Phys. Rev. Lett. {\bf 45}, 2084.

\bibitem{BV_el} Beacom J. F. and Vogel P., 1999, Phys. Rev. Lett.
{\bf 83}, 5222.

\bibitem{pv_osc_1} Boehm F. {\it et al.}, 2000a, Phys. Rev. Lett. {\bf 84}, 3764.

\bibitem{pv_osc_2} Boehm F. {\it et al.}, 2000b, Phys. Rev. D{\bf 62}, 072002.

\bibitem{pv_osc_3} Boehm F. {\it et al.}, 2000c, Phys. Rev. D{\bf 62}, 092005.

\bibitem{pv_osc_4} Boehm F. {\it et al.}, 2001, hep-ex/0107009, Phys. Rev. D, submitted.

\bibitem{past-reactors}  Boehm F. and Vogel P.,  1992
{\it Physics of Massive Neutrinos}, (Cambridge Univ. Press, Cambridge).

\bibitem{felix} Boehm F., 2001, chapter in {\it Current aspects of Neutrino Physics},
D. Caldwell ed., (Springer, Heidelberg), to be published.

\bibitem{cp1}Cabibbo N., 1978, Phys. Lett. {\bf 72B}, 333.

\bibitem{Danby} Danby G. {\it et al}, 1962, Phys. Rev. Lett. {\bf 9}, 36.

\bibitem{fiss_calc} Davis B. R. {\it et al.}, 1979,  Phys. Rev. C{\bf 19}, 2259.
                   

\bibitem{Declais} Declais Y. et al, 1994, Phys. Lett. {\bf B338}, 383.

\bibitem{Derbin} Derbin A. I. {\it et al.}, 1993, JETP Lett. {\bf 57}, 768.

\bibitem{Detwiler} Detwiler J., 2000, \textsc{Kamland} note 00-06, unpublished.

\bibitem{Eadie} Eadie W.~T.{\it et al.}, 1971, 
{\it Statistical methods of experimental physics}, (North-Holland, Amsterdam).

\bibitem{Eders} Eders G., 1966, Nucl. Phys. {\bf 78}, 657.

\bibitem{} Eitel K., 2000, Nucl. Phys. Proc. Suppl. {\bf 91}, 191.

\bibitem{Fayans1} Fayans S. A., 1985, Sov. J. Nucl. Phys. {\bf 42}, 590.

\bibitem{feldman} Feldman G.J. and Cousins R.D., 1998, Phys. Rev. D{\bf 57}, 3873.

\bibitem{Zuber-Fisher}  Fisher P., Kayser B. and McFarland K.S., 1999,
Ann. Rev. Nucl. Part. Sci. {\bf 49}, 481.

\bibitem{} Fukuda S.  {\it et al.}, 2000, Phys. Rev. Lett.
{\bf 85}, 3999.

\bibitem{} Fukuda S. {\it et al.}, 2001, preprints hep-ex/0103032 and 0103033,
submitted to Phys. Rev. Lett.

\bibitem{SK_el} Fukuda Y. {\it et al}., 1999, Phys. Rev. Lett. {\bf 82}, 2430.


\bibitem{Fukuda} Fukuda Y. {\it et al.}, 1998, Phys. Rev. Lett.
{\bf 81}, 1562.

\bibitem{trigger} Gratta G. {\it et al.}, 1997,
Nucl. Instr. and Meth. {\bf A400}, 456.

\bibitem{RPP} Groom  D.E. {\it et al.},  2000, Europ. Phys. J. C{\bf 15}, 1.

\bibitem{Hahn:1989zr} Hahn A.A. {\it et al.}, 1989, Phys. Lett. {\bf B218}, 365.

\bibitem{perkins} Harrison P.F., Perkins D.H. and Scott W.G., 1996, Phys. Lett. 
{\bf B349}, 137.

\bibitem{MINUIT} James F., 1994, MINUIT Reference Manual, vers. 94.1.

\bibitem{atm} Kajita T. and Totsuka Y.,  2001, Rev. Mod. Phys. {\bf 73}, 85.

\bibitem{solar} Kirsten T. A., 2000,  Nucl. Phys. Proc. Suppl.
{\bf 87}, 152.

\bibitem{fiss_klap} Klapdor  H-V. and Metzinger J., 1982a, Phys. Rev. Lett. {\bf 48}, 127.

\bibitem{} Klapdor H-V.  and Metzinger J., 1982b, Phys. Lett. {\bf B112}, 22.

\bibitem{nutau} Kodama K. {\it et al.}, 2001, Phys. Lett. {\bf B504}, 218.

\bibitem{fiss_russ} Kopeikin V. I., Mikaelyan L. A., and Sinev V. V., 1997,
Phys. Atom. Nucl. {\bf 60}, 172.

\bibitem{Kozlov} Y. V. Kozlov Y. V. {\it et al.}, 2000, Phys. Atom. Nucl. {\bf 63}, 1016.

\bibitem{Krauss} Krauss L. {\it et al.}, 1984, Nature {\bf 310}, 191.

\bibitem{ill} Kwon H. {\it et al.}, 1981, Phys. Rev. D{\bf 24}, 1097.

\bibitem{LS} Llewellyn-Smith C. H., 1972, {\it Phys. Rep.} {\bf 3}, 261.

\bibitem{Maki} Maki Z., Nakagawa M., and Sakata S., 1962,
Progr. Theor. Phys. {\bf 28}, 870.

\bibitem{earth_antinu} Marx G., 1969, Czech. J. of Phys. B{\bf 19}, 1471.

\bibitem{}Marx G. and Lux I., 1970, Acta Phys. Acad. Hung. {\bf 28}, 63.


\bibitem{krasnoyarsk} Mikaelyan L., 2000,
Nucl. Phys. Proc. Suppl. {\bf 91}, 120.

\bibitem{MSW3} Mikheyev S. P. and Smirnov  A. Yu., 1986a,
Sov. J. Nucl. Phys. {\bf 42}, 913. 

\bibitem{MSW3} Mikheyev S. P. and Smirnov  A. Yu., 1986b, Nuovo Cimento {\bf 9C}, 17.


\bibitem{Lester_thesis} Miller L., 2000, PhD thesis, (Stanford
University).

\bibitem{chooz_osc_3} Nicol\`o D.,  1999, Thesis, Publication
of Scuola Normale Superiore, Pisa.

\bibitem{Pasierb} Pasierb E. {\it et al.}, 1979, Phys. Rev. Lett.
{\bf 43}, 96. 
\bibitem{Perl} Perl  M. L. {\it et al.}, 1975,  Phys. Rev. Lett. {\bf 35}, 1489. 

\bibitem{scint} Piepke A. {\it et al.}, 1999, Nucl. Instr. and Meth. 
{\bf A432}, 392.

\bibitem{Pontecorvo} Pontecorvo B., 1958, Sov. Phys. JETP {\bf 6},
429.

\bibitem{Pontecorvo2} Pontecorvo B., 1967, Zh. Eksp. Theor. Phys. {\bf 53}, 1717.

\bibitem{SK3} Okamura K., 1999, Ph.D. Thesis, (University of Tokyo).

\bibitem{Raghavan_terr} Raghavan R.S. {\it et al.}, 1998, Phys. Rev. Lett. {\bf 80}, 635.

\bibitem{Reines} Reines F. and C. L. Cowan C. L., Jr., 1953, 
Phys. Rev. {\bf 92}, 830. 

\bibitem{Reines2} Reines F. and C. L. Cowan C. L., Jr., 1959,
Phys. Rev. {\bf 113}, 273.

\bibitem{Reines_el} Reines F., Gurr H. S. , and Sobel H. W., 1976,
 Phys. Rev. Lett. {\bf 37}, 315.

\bibitem{} Reines F.  {\it et al.}, 1980, Phys. Rev. Lett.
{\bf 45}, 1307.


\bibitem{Riley} Riley  S. P. {\it et al.}, 1999, Phys. Rev. C {\bf 59}, 1780.


\bibitem{Schreckenbach:1985ep} Schreckenbach K. {\it et al.}, 1985, Phys. Lett. 
{\bf B160},325.

\bibitem{Tengblad} Tengblad O. {\it et al.}, 1989, Nucl. Phys. {\bf A503}, 136.

\bibitem{krasnoyarsk1} Vidyakin G. S. {\it et al.}, 1994, JETP {\bf 59}, 390.

\bibitem{} Vogel P. {\it et al.}, 1981, Phys. Rev. C{\bf 24}, 1543.

\bibitem{Vogel} Vogel P., 1984, Phys. Rev. D{\bf 29}, 1918.


\bibitem{VE} Vogel P. and Engel J., 1989, Phys. Rev. D{\bf 39}, 3378.


\bibitem{BV_ang} Vogel P. and Beacom J. F., 1999,  Phys. Rev. D
{\bf 60}, 053003.

\bibitem{swap} Wang  Y-F.{\it et al.}, 2000, Phys. Rev. D{\bf 62}, 013012.

\bibitem{} Wang  Y-F.{\it et al.}, 2001, Phys. Rev. D{\bf 62}, 013012.  

\bibitem{Wilkinson}
Wilkinson D.H., 1982, {\it Nucl. Phys.} {\bf A377}, 474.

\bibitem{winter_book} Winter K. ed., 1991,  {\it Neutrino Physics}, 
(Cambridge Univ. Press, Cambridge).


\bibitem{Wojcicki} Wojcicki S., 2001a, Nucl. Phys. Proc. Suppl, {\bf 91}, 216.
 
\bibitem{} Wojcicki S., 2001b, talk  at {\it Neutrino Telescopes},
Venice, March 2001, to be published in proceedings.

\bibitem{MSW} Wolfenstein L., 1979, Phys. Rev. D{\bf 17}, 2369.

\bibitem{MSW2} Wolfenstein L., 1980, Phys. Rev. D{\bf 20}, 2634. 

\bibitem{goesgent}Zacek G.,  1984, Dr. Thesis, (Technical University Munich).

\bibitem{goesgen} Zacek G. {\it et al.}, 1986, Phys. Rev. D{\bf 34}, 2621.

\bibitem{} Zuber K., 1998, Physics Reports {\bf 305}, 296.

\end{thebibliography}
\end{document}